\documentstyle[epsf]{mn}

\newcommand{\hb}{\ifmmode {\rm H}\beta \else H$\beta$\fi}
\newcommand{\lya}{\ifmmode {\rm Ly}\alpha \else Ly$\alpha$\fi}
\newcommand{\ciii}{\ifmmode {\rm C}\,{\sc iii} \else C\,{\sc iii}\fi}
\newcommand{\civ}{\ifmmode {\rm C}\,{\sc iv} \else C\,{\sc iv}\fi}
\newcommand{\mgii}{Mg\,{\sc ii}}
\newcommand{\nv}{N$\,{\sc v}$}
\newcommand{\approxlt}{\raisebox{-.5ex}{$\;\stackrel{<}{\sim}\;$}}

\begin{document}
\title{ Quasar Tomography:
	Unification of Echo Mapping and Photoionisation Models}
\author[Keith Horne et~al.]
{ 
Keith~Horne$^1$,
Kirk~T.~Korista$^2$,
Michael~R.~Goad$^3$
\\
$^1$Physics \& Astronomy, St.~Andrews University,
	North Haugh, St.\ Andrews, KY16 9SS, Scotland, UK.
	\verb+kdh1@st-and.ac.uk+.
\\
$^2$Department of Physics, Western Michigan University,
	1120 Everett Tower, Kalamazoo, MI, 49008, USA.
	\verb+korista@wmich.edu+.
\\
$^3$Physics \& Astronomy, University of Southampton,
	Southampton SO17~1BJ, UK.
	\verb+mrg@astro.soton.ac.uk+.
}

\date{Accepted 2002 Sep 4, Submitted 2002 Jul 26 }
 
\pagerange{\pageref{firstpage}--\pageref{lastpage}} \pubyear{????}

\maketitle

\label{firstpage}

\begin{abstract}

Reverberation mapping aims to use time-delayed variations in
photoionised emission lines to map the  geometry and kinematics of 
emission-line gas in the vicinity of an active galactic nucleus.
By fitting to a time-variable emission line profile,
one can reconstruct a 2-dimensional map
$\Psi(\tau,v)$, where $\tau$ is the time delay and $v$
is the Doppler shift, for each emission line.
In this paper we develop quasar tomography, which combines
the time delay and velocity information with
photoionisation physics in order to map the
reprocessing region using information assembled from many
different emission lines.
The observed spectral variations are modelled
in terms of direct light from the active nucleus
and time-delayed reprocessed light from surrounding gas clouds.  
We use the photoionisation code CLOUDY to evaluate
line and continuum reprocessing efficiencies $\epsilon( \lambda,
\Phi_{\sc H}, n_{\sc H}, N_{\sc H}, \theta )$ for clouds of 
hydrogen density $n_{\sc H}$ and column density $N_{\sc H}$ 
exposed to hydrogen-ionising photon flux $\Phi_{\sc H}$.  
The gas distribution is described by a 5-dimensional map,
the differential covering fraction 
$f( R, \theta, n_{\sc H}, N_{\sc H}, v )$,
which we reconstruct from the 2-dimensional data
$F_\lambda( \lambda, t )$ by using maximum entropy techniques.
Tests with simulated data and a variety of geometries
(shells, rings, disks, clouds, jets)
are presented to illustrate some of the capabilities
and limitations of the method.
Specifically, we reconstruct 3-dimensional geometry-density maps,
$f(R, \theta, n_{\sc H})$,
by fitting to well-sampled light curves for the continuum
and 7 ultraviolet emission lines.
The maps are distorted in ways that we understand and discuss.
The most successful test recovers a hollow shell geometry,
determining correctly its radius and density.
The data constrain the ionisation parameter
$U\propto R^{-2}n_{\sc H}^{-1}$ to about 0.1~dex,
the radius $R$ to 0.15~dex,
and $n_{\sc H}$ to 0.3~dex.
We expect better constraints to arise from future fits using
more lines and velocity profiles as well as velocity-integrated
line fluxes.
The maps are sensitive to the assumed distance,
offering some prospects for using
emission-line reverberations to measure the
luminosities and distances of active galactic nuclei.

\end{abstract}

\section{Introduction}
\label{sec:intro}

\subsection{ Photoionisation Modelling }

Our understanding of the emission-line spectra
of active galactic nuclei (AGNs) has progressed considerably
since the late 1970s,
when photoionisation calculations were first used 
to model observed spectra in an effort to
determine the physical characteristics of the emission-line gas 
(e.g., Davidson \& Netzer 1979).
Sophisticated photoionisation codes now embody
our current detailed knowledge and understanding of atomic physics
and physical processes in astrophysical plasmas.
The evolution of these codes toward increasing realism is
guided by detailed comparison of observed and predicted spectra,
made possible by advances in high-speed computing.

In the photoionisation code CLOUDY (Ferland et al. 1998), a prescribed
spectrum $L_\lambda$ is directed into a gas cloud at radius $R$
characterized by its hydrogen density $n_{\sc H}$ and hydrogen column
density $N_{\sc H}$.  The temperature and ionisation state of the gas
in successive zones inside the cloud are evaluated, as well as the
line and continuum fluxes that emerge from the front and back faces of
the cloud.
For given elemental abundances,
the calculated emission-line flux ratios depend upon
$L_\lambda$, $R$, $n_{\sc H}$, and $N_{\sc H}$.
However, the primary dependence is upon the ionisation parameter
\begin{equation}
U \equiv
 \frac{Q_{\sc H}}{4 \pi R^2 n_{\sc H} c}
	 = \frac{\Phi_{\sc H}}{n_{\sc H} c}\ ,
\end{equation}
where
\begin{equation}
	Q_{\sc H} = \int_0^{\lambda_{\sc H}} 
		\frac{\lambda L_\lambda d\lambda} { h c } 
\end{equation} 
is the rate at which the source emits hydrogen-ionising photons
$(\lambda < \lambda_{\sc H}=912$\AA),
and
\begin{equation}
	\Phi_{\sc H} = \frac{Q_{\sc H}}{4 \pi R^2}
\end{equation} 
is the flux of hydrogen-ionising photons incident on the cloud.
By comparing observed emission-line ratios
(which are rather similar for most objects)
with predictions from single-cloud photoionisation calculations,
``typical'' physical conditions in the emission-line gas are
inferred to be
$n_{\sc H}\sim10^{9.5}$cm$^{-3}$,
$U\sim10^{-2}$,
$\Phi_{\sc H}\sim10^{18}$cm$^{-2}$s$^{-1}$
(Davidson \& Netzer 1979).
For these parameters the distance of the gas cloud from the nucleus is 
\begin{equation}
\label{eqn:photorad}
	R = \left( \frac{Q_{\sc H} } {4\pi \Phi_{\sc H} }\right)^{1/2}
	\sim 100 L_{44}^{1/2}~{\rm light~days}\ ,
\end{equation}
where $L_{44} \sim Q_{\sc H}\ h c / \lambda_{\sc H}$ is
the hydrogen-ionising luminosity of the source in
units of $10^{44}$~erg~s$^{-1}$.

Since the derived gas temperatures are low  ($T \sim 10^{4}$~K),
the broad emission-line profiles ($V_{\sc FWHM} \sim 5000$~km~s$^{-1}$ )
are interpreted as arising
from bulk motion of the emission-line gas, most likely arising
from gravitational acceleration of the gas by the nucleus.
A rough virial mass may then be estimated by combining the 
$V_{\sc FWHM}$
with the radius estimated above,
\begin{equation} M \sim 4\times 10^{8} M_\odot
	\left( \frac{V_{\sc FWHM}}{5000~{\rm km~s}^{-1} } \right)^2
	\left( \frac{R}{ 100~{\rm light~ days} } \right)\ .
\end{equation}

The rough similarity of observed emission-line ratios
in high-luminosity quasars and lower-luminosity Seyfert galaxies
suggests that although these active galactic nuclei
span a wide range of luminosity,
their emission-line regions are nevertheless characterized by
roughly the same ionisation parameter.
Moreover, in many cases comparison of velocity widths and profiles of
different emission lines suggests that $U$ is roughly independent
of $R$ and hence $n_{\sc H} \propto R^{-2}$.
These regularities in the observed properties of quasar emission lines
have become recognized as a problem because it is not known what
mechanisms may regulate conditions in the emission-line gas to ensure
a constant ionisation parameter.

A plausible solution to this ``fine-tuning'' problem has recently emerged.
The LOC model (Locally Optimally-emitting Clouds, 
e.g., Baldwin et~al. 1995 )
proposes that the photoionised gas is not adequately
characterized by a single density and column density at each radius,
	$n_{\sc H}(R)$, $N_{\sc H}(R)$.
Instead, a whole population of gas clouds,
	$f(R,n_{\sc H},N_{\sc H})$,
provides for a wide range of $n_{\sc H}$ and $N_{\sc H}$ at each $R$.
The total emission in a given line
is then obtained by summing contributions from all these different
cloud types, but most of the emission arises from the specific cloud types
that are maximally efficient in producing that particular line.
The photoionisation models indicate that high reprocessing efficiency
in a given line is achieved by a sub-population of the clouds
defined primarily by a specific range of ionisation parameter.
In the LOC model, the optimal conditions are obtained
at many radii in a given source by employing
lower density gas when farther from the nucleus,
and over a wide range of source luminosity
by using gas at larger radius in higher-luminosity sources.
Thus the apparent fine tuning is a consequence of photoionisation
physics rather than specific conditions in the emission-line gas.

\subsection{ Reverberation Mapping }

The assumption that the emission lines are driven by photoionisation
is amply validated by variability monitoring campaigns. 
The active nuclei in half a dozen Seyfert~1 galaxies have been
studied by the AGN~Watch collaboration (Alloin et al. 1994),
combining {\em IUE} and ground-based observations
to monitor variability in the ultraviolet and optical spectra.
The AGN~Watch campaigns extend over many months, with time sampling
typically a few days (Peterson 1993). 
In all cases the observed variations in the continuum flux, $F_c(t)$,
are accompanied by correlated variations in emission-line
fluxes, $F_\ell(t)$.
Moreover, the emission line variations lag
behind those seen in the continuum.
The time delay, $\tau$, can be
interpreted as light-travel time.
This effectively measures the radial distance,
\begin{equation}
	R \sim \frac{ c \tau }{(1 + z )}
\ ,
\end{equation}
from the nucleus
(more accurately from the continuum-emitting regions close to the
nucleus) to the line-emitting region.
Here $z$ is the cosmological redshift of the source.

The time delays are usually estimated
by cross-correlating the observed line and continuum light curves
(e.g., Edelson \& Krolik 1988, Koratkar \& Gaskell 1991
White \& Peterson 1994).
These results have revealed time delays of days to weeks,
with shorter delays in the higher ionisation lines,
consistent with more highly ionised gas occurring closer to the nucleus 
(e.g., Clavel et~al. 1991). 

More detailed information can be extracted by
``reverberation mapping'' (Bahcall, Kozlovsky \& Salpeter 1972,
Blandford \& McKee 1982).
A delay map $\Psi(\tau)$ can be recovered by fitting
to high-quality densely-sampled light curves,
for example by using the linear echo model
\begin{equation} 
	F_\ell(t) = \int_0^{\tau_{max}}
		\Psi(\tau)\ F_c(t-\tau)\ d\tau\ .
\end{equation}
Maximum entropy fitting techniques have proven to be useful
in this regard
(e.g., Horne, Welsh \& Peterson 1991, Krolik et al. 1991, Horne 1994),
as have regularized linear inversion methods
(Vio, Horne \& Wamsteker 1994, Pijpers \& Wanders 1994, Krolik \& Done 1995).

For the strongest lines, \civ\ in particular,
data quality can be sufficient to record reverberation effects
in the velocity profiles, allowing the recovery of
kinematic information in the form of a velocity--delay map $\Psi(v,\tau)$
(e.g., Wanders et al. 1995, Ulrich \& Horne 1996, Done \& Krolik 1996).
Results to date show that the red and blue wings of the \civ\
emission line respond together with smaller delays than the
line centre.
Models involving purely radial inward or outward motions are
ruled out, but a variety of models involving primarily
random or azimuthal motions, as in a disk, are permitted.

Time delays measured from \hb\ and optical continuum light curves
of AGNs with a range of luminosities
($10^{41.8}$~erg~s$^{-1}
< \lambda L_\lambda({\rm 5100\AA})
< 10^{45.7}$~erg~s$^{-1}$)
establish an empirical relationship of the form
\begin{equation}
	R_{\hb} \sim 30\ 
\left( \frac{ \lambda L_\lambda({\rm 5100\AA}) }
	{ 10^{44} {\rm erg~s}^{-1} }
	\right)^{0.7} {\rm light~ days}
\end{equation}
(Kaspi et al. 2000).
This is consistent with the photoionised region
increasing in size as the ionising luminosity increases.

The echo-mapping experiments have forced us to realize that
emission-line gas is present much closer to the nucleus than the
radius estimated in Equation~(\ref{eqn:photorad})
on the basis of single-cloud photoionisation models.
For a given $U$, the smaller radius requires a higher density,
$n_{\sc H} \sim 10^{11}$cm$^{-3}$ 
(Ferland et~al. 1992).
In earlier work,  the \ciii]/\civ\ ratio was used to argue
against such high densities.
This argument may be incorrect, however,
if \ciii] and \civ\ form in different regions with different 
densities, as is suggested in several cases
where the \ciii] lag is longer than the \civ\ lag
(Clavel et~al. 1991, Reichert et~al. 1994).
The smaller radii also reduce the inferred virial masses.

\subsection{ Quasars as Cosmological Probes }

Quasars are potentially important as cosmological probes.
Their high luminosity renders them visible at large redshifts.  
However, quasars are poor standard candles,
with luminosities ranging over many orders of magnitude.  
In low-redshift samples the emission-line ratios and
equivalent widths are correlated with luminosity
(Baldwin 1977; Baldwin et~al. 1978; Kinney et~al. 1990),
but the dispersion is too large to be of much use
in estimating luminosities or distances.
Separating luminosity and evolution effects is also difficult
with magnitude-limited samples that tend to include
high-luminosity objects at high redshift and low-luminosity objects at
low redshift.

Distances may alternatively be estimated by using light travel time delays
from reverberation effects to measure the linear size
of something whose angular size can be directly observed or
inferred from the observed spectrum.
For example, the distance to supernova 1987A was estimated from
time-delayed enhancements in photoionised emission lines
produced when the ultraviolet flash from the explosion
first reached the inclined circumstellar ring that is
resolved by HST (Panagia et al.~1991).
Can we apply a similar method to AGNs?

For AGNs a method based on reverberations in continuum emission 
from the irradiated surface of the accretion disc
has been demonstrated by Collier et al.~(1999).
A steady-state disc is predicted to have a characteristic
temperature profile of the form $T \propto R^{-3/4}$.
With $\tau \sim (1+z)R/c$ from light travel time,
and $kT \sim hc(1+z)/4\lambda$ from blackbody radiation,
the temperature decreasing with radius implies
a time delay increasing with wavelength as
$\tau \propto \lambda^{4/3} (1+z)^{-1/3}$.
Just such an effect was found, with delays of order 1-2 days
between ultraviolet and optical continuum variations
in the Seyfert 1 galaxy NGC~7469.
The resulting redshift-independent distance yields
$H_0\sqrt{\cos{i}/0.7} = 42\pm8$~km~s$^{-1}$Mpc$^{-1}$,
where $i$ is the inclination of the disk axis to the line of sight,
expected to be less than $60^\circ$ for Seyfert 1 galaxies.

In \S\ref{sec:cosmology} of this paper
we discuss a new method of estimating distances to AGNs
based on reverberation in the photoionised emission lines.
The basic idea is to require the radii of photoionised regions
estimated from time delays 
to match the radii derived from comparison of photoionisation
calculations with the observed emission-line ratios.
This may provide a new direct method of determining
distances, based on a straightforward interpretation of
time delays and photoionisation physics.

\subsection{ Quasar Tomography }

The goal of quasar tomography, as developed in this paper,
is to unify echo mapping and photoionisation modelling.
Echo mapping reveals the sizes of emission-line regions
while photoionisation modelling uses emission-line flux ratios
to determine physical conditions within the emission-line gas.
Current echo mapping techniques define delay maps for each line
independently of the other lines and without regard
to photoionisation physics.
More complete information can be extracted
from high-quality time-resolved emission-line spectra
by combining these two approaches.

Following the spirit of the LOC model,
we offer the emission-line gas freedom to fill out a
very general 5-dimensional map,
$f( R, \theta, n_{\sc H}, N_{\sc H}, v)$,
giving the differential covering fraction of the gas clouds.
We then expect observations of time-dependent spectra 
$F_\lambda( \lambda, t )$
to reveal the geometry, physical conditions, and
kinematics of the gas
by placing constraints that define the structure of the 5-D cloud map.

We are aiming to fit simultaneously,
in far greater detail than has hitherto been attempted,
the time-dependent fluxes and velocity profiles
of numerous emission lines recorded in high quality spectra.
Our methods and assumptions in modelling the reverberating spectrum
$F_\lambda( \lambda, t )$ are developed in Appendix~\ref{app:forward}.
Appendix~\ref{app:inverse} then discusses 
the maximum entropy techniques we employ to
recover the cloud map $f( R, \theta, n_{\sc H}, N_{\sc H}, v )$
by fitting to the observations of $F_\lambda( \lambda, t )$.

In \S\ref{sec:cloudmaps} 
we illustrate some of the capabilities and
limitations of quasar tomography by presenting the results of test
reconstructions from simulated datasets.  A variety of possible
geometries is considered.
In \S\ref{sec:cosmology} we assess prospects
for using this approach to derive redshift-independent distances and
luminosities.  Concluding remarks are made in \S\ref{sec:summary}.

\section{ Cloud Maps recovered from Simulated Datasets }
\label{sec:cloudmaps}

\subsection{ Appearance of Different Geometries }
\label{sec:geometries}

It is important to realize that most types of observational data
obtained from a distant unresolved object are unaffected if the object
is rotated by an angle $\phi$ around the line of sight.
A possible exception is polarimetric data, where the position angle
of linearly polarized light is measured.
In quasar tomography we employ information from time delays
and emission-line ratios that are sensitive to $R$ and $\theta$
but not to $\phi$.
Thus we should not expect to be able to recover the 3-dimensional
source geometry, $f(R,\theta,\phi)$,
but rather the 2-dimensional geometry,
\begin{equation}
	f(R,\theta) = \int f(R,\theta,\phi)\ d\phi
\ ,
\end{equation}
that arises by rotating the 3-dimensional geometry
around the line of sight to the observer.

\begin{figure}
\vspace*{10.5cm}
\begin{picture}(8,11)
\epsfxsize=8cm
\epsfysize=11cm
 \makebox[8cm][l]{\epsfbox{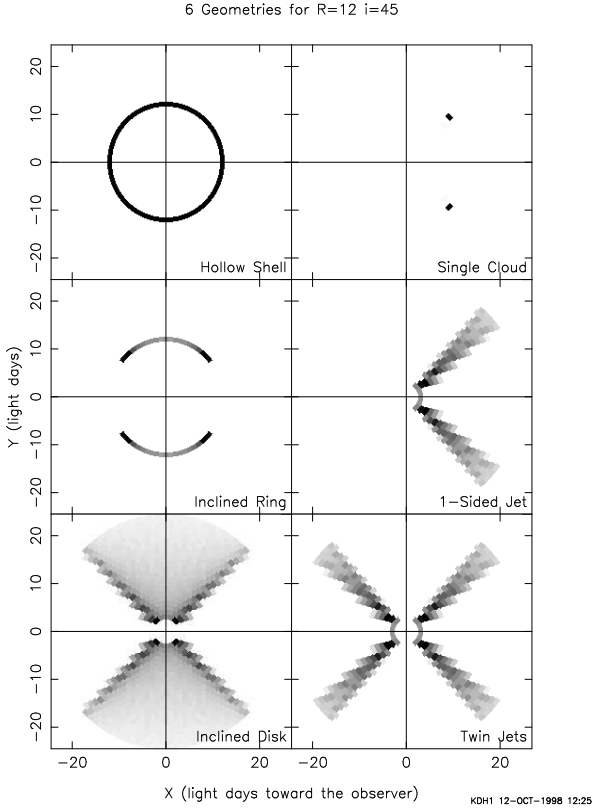}}
\end{picture}\par
\caption[]
{\small
Maps showing the appearance of six different geometries
after being rotated around the line of sight (X axis).
\label{fig:geometries}}
\end{figure}

To illustrate the effect of this ambiguity,
Fig.~\ref{fig:geometries} displays the appearance of several simple
geometries after projection around the line of sight onto the $X,Y$ plane.
Here $X = -R \cos{\theta}$ is measured from the nucleus towards the observer,
whose location is off to the right-hand side of the figure,
and $Y = R \sin{\theta}$ is measured perpendicular to the line of sight.
The ambiguity in $\phi$ means that we can project any given geometry
by rotating it around the line of sight.  This yields a projected
2-dimensional geometry that is symmetric to reflection
through the $X$ axis.
A single cloud (upper-right panel) projects to
a spot at the appropriate radius $R$ and azimuth $\theta$,
plus its mirror image reflected through the $X$ axis.
A hollow spherical shell (upper-left panel)
projects to a complete ring with radius $R$ equal to that of the shell.

An inclined ring or annulus (middle-left panel)
projects to an arc of radius $R$ that straddles the $+Y$ axis,
plus its mirror image on the $-Y$ axis.
The arc spans $\theta = 90^\circ \pm i$,
where $i$ is the inclination angle between the ring axis and the line of sight.
The figure shows the projection of a ring inclined by $i=45^\circ$.
If we tilt the ring axis toward the observer, the two arcs become
shorter, shrinking to a pair of spots at $Y=\pm R$ as $i\rightarrow0$.
Tilting the ring axis away from the observer makes the arcs longer
until their endpoints meet to form a complete ring as $i\rightarrow90^\circ$.
This edge-on ring may still be distinguished from the hollow shell,
however, because a relatively large fraction of the ring's
circumference projects to the region near the endpoints of the arc.

An inclined disk (lower-left panel) is assembled from a set
of co-axial annuli with different radii.
The disk annuli project to arcs of different radii that fill out
the sector bounded by the smallest and largest arcs.
This makes an X-shaped pattern with the sectors above and below
the X filled in and empty sectors to the sides.
As with the inclined ring, tilting the disk toward (away from)
the line of sight closes (opens) the filled sector of the X.

A 1-sided hollow conical jet (middle-right panel) is easily recognizable
from its projection to a cone emerging from the nucleus.
The jet inclination $i$ is the angle between the cone and the $X$ axis,
and the jet opening angle $\omega$ is the same as that of the cone.
Of course the mirror image of the jet, projected across the $X$ axis,
is also present after rotation about the line of sight.
If we increase the jet opening angle, the corresponding cone
becomes wider, eventually overlapping with its mirror image when
$\omega > i$.
Finally, twin jets extending in opposite directions (lower-right panel)
project to an X-shaped pattern of cones.

The main point we wish to make here is that these six geometries,
and many others, have distinct appearences after projection
by rotation around the line of sight onto the $X-Y$ plane
that we aim to recover from the data.  
In \S\ref{sec:fitslc} we present and discuss maps of these same
six geometries reconstructed from simulated datasets.

\subsection{ Synthetic Light Curves }
\label{sec:slc}

Fig.~\ref{fig:shell60tru_fit} shows a simulated dataset
calculated using the methods outlined in Appendix~\ref{app:forward}
for the geometrically-thin spherical shell shown in Fig~\ref{fig:geometries}.  
All clouds within the shell are assumed to have a gas
hydrogen density of $n_{\sc H}=10^{11}$cm$^{-3}$, and a hydrogen column
density of $N_{\sc H}=10^{23}$cm$^{-2}$.

\begin{figure}
\vspace*{10.5cm}
\begin{picture}(8,11)
\epsfxsize=8cm
\epsfysize=11cm
 \makebox[8cm][l]{\epsfbox{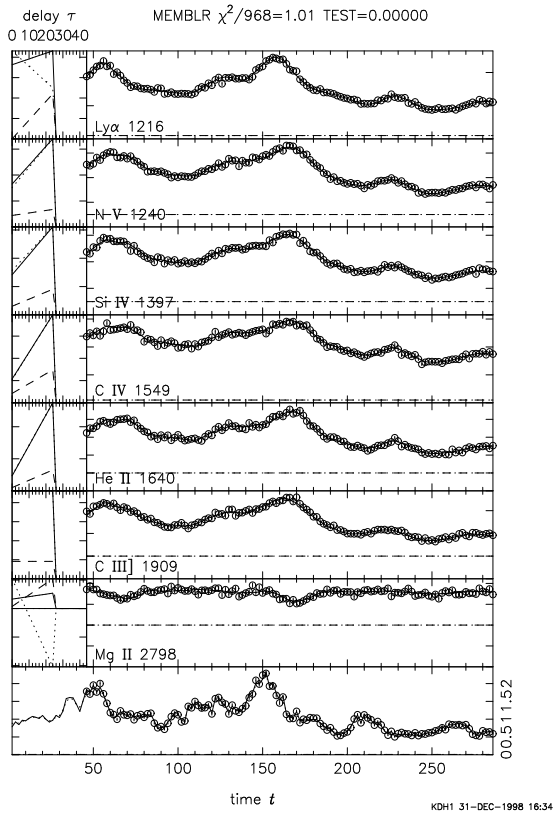}}
\end{picture}\par
\caption[]
{\small
Simulated continuum and emission-line light curves 
for a thin spherical shell.
The continuum light curve (lower panel)
convolved with the appropriate delay map (left panels)
gives the corresponding emission-line light curve (right panels).
The delay maps account for light travel time delays
as well as non-linear and anisotropic emission-line responses
to variations in the ionising radiation.
See text \S\ref{sec:slc} for details.
\label{fig:shell60tru_fit}}
\end{figure}

The bottom panel shows the driving light curve, $F_c(t)$,
and the responding
emission-line light curves $F_\ell(t)$
are shown for 7 different ultraviolet emission lines in the 7 panels above.
The data points in these panels were generated by first evaluating the true
light curve, and then adding a Gaussian random variate with zero mean
and standard deviation as shown by the vertical error bars 
(smaller than the data points plotted)
to simulate 3\% observational errors.
The horizontal dashed lines show the background fluxes
$F_B(\lambda)$
for each line, included in the model to represent constant
line emission that is not driven by photoionisation.
Light curves similar to these would be 
possible to obtain in one year of observation
by observing once every two days.

Delay maps $\Psi_\ell(\tau)$ are shown in Fig.~\ref{fig:shell60tru_fit}
to the left of the corresponding light curve for each emission line 
$\ell$.
The delay maps for this geometrically-thin spherical shell with $R=12$
light days extend from 0 to $2R/c=24$ light days.  
The prompt response (at $\tau = 0$) arises from the near edge of the
shell, while the response with maximum delay ($\tau = 2R(1+z)/c$)
arises from the far edge of the shell.
When the line emission is isotropic (e.g., \ciii] in 
Fig.~\ref{fig:shell60tru_fit})
then the delay map for the spherical shell is flat-topped,
with equal response in each interval of $\tau$ from 0 to $2R(1+z)/c$.
For the other lines the delay maps have a positive slope
(response increasing with increasing $\tau$)
due to their inward anisotropy
(enhanced emission at large delays arising from the far edge of the
shell).

The exact angular pattern of anisotropic line radiation depends
of course on details of geometry and radiative transfer in
the reprocessing region.
These details are beyond the scope of current 1-dimensional
photoionisation codes.
We are therefore handling the anisotropic line emission by
using CLOUDY to evaluate the inward and outward emission
of the line,  and then interpolating linearly in $\cos{\theta}$
to obtain the emissivity in direction $\theta$.
For further discussion see \S\ref{sec:aim}.

The driving light curve 
(bottom panel of Fig.~\ref{fig:shell60tru_fit})
varies by a factor $\sim 4$ in this example,
and the delay maps can change appreciably when the ionising flux
changes by such a large factor between the bright and faint states.
To illustrate this,
the left panels of Fig.~\ref{fig:shell60tru_fit} give
delay maps corresponding to both the bright state
(solid curves) and the faint state (dashed curves)
corresponding to the maximum and minimum of the driving light curve.
A third delay map (dotted curve) is the difference
between the bright and faint state delay maps,
scaled to have the same peak as the bright state map.
This presentation makes it easy to spot non-linear line responses.
The scaled difference map (dotted curve) will differ from the
bright state map (solid curve)
only if there is a significant non-linear response in that emission line.
It can be seen that the responses in all lines
are positive and essentially linear,
with the exceptions of \lya, and \mgii.

\lya\ shows a marked non-linear response primarily due to a change in the
line radiation pattern between the faint and bright states. 
In the faint state, \lya\ has a strong inward anisotropy
because its large optical depth causes the bulk of
its emission to emerge from the irradiated faces of clouds
(Ferland et~al. 1992; O'Brien et~al. 1994).
Thus we see stronger \lya\ emission from clouds on the far side
of the shell than we do from those on the near side.
In contrast, \lya\ emission is almost isotropic
in the bright state because the hydrogen-ionisation front pushes
through the clouds, so that those on the near side of the shell
become visible.
Thus in the top left panel of Fig.~\ref{fig:shell60tru_fit}
we see that in the low state (dashed delay map)
the prompt \lya\ response at $\tau = 0$,
arising from the near side of the shell,
is essentially zero.
In the high state (solid delay map),
the \lya\ response is almost independent of $\tau$,
indicating nearly isotropic \lya\ emission from the spherical shell.

The \mgii\ response is also non-linear.
Moreover, the \mgii\ response on the far side of the shell is negative,
meaning that \mgii\ emission decreases with increasing ionising flux.
In the faint state, the \mgii\ radiation is dominated by
emission from the irradiated sides of the clouds,
producing a strong inward anisotropy with very low prompt response.
In the bright state,
the \mgii\ emitting zone becomes significantly depleted as the
hydrogen-ionisation front moves deeper into the cloud, and the line
emerges more isotropically. The depletion of the partially ionised
zone where this line is formed results in a net negative response for
this line (see also Goad et~al. 1993; O'Brien et~al. 1995). 
The negative response of \mgii\ can be seen directly in the light curves,
where the \mgii\ flux has minima at times when the other line
fluxes have maxima.

In the spherical shell example we are considering,
diagnosis of the delay maps reveals isotropic and anisotropic emission,
linear and non-linear responses,
and positive and negative responses in the various emission lines.
Each of these effects conveys information about the possible
location and physical state of the gas clouds that are
responding to changes in the flux of ionising radiation
represented in the dataset by the driving light curve.
The information we seek is not readily available, however,
as it appears in the delay maps rather than in the observed light curves.
We must therefore consider to what extent it is possible to
recover the geometry and physical conditions
by fitting to the light curves.

\begin{figure}
\vspace*{10.5cm}
\begin{picture}(8,11)
\epsfxsize=8cm
\epsfysize=11cm
 \makebox[8cm][l]{\epsfbox{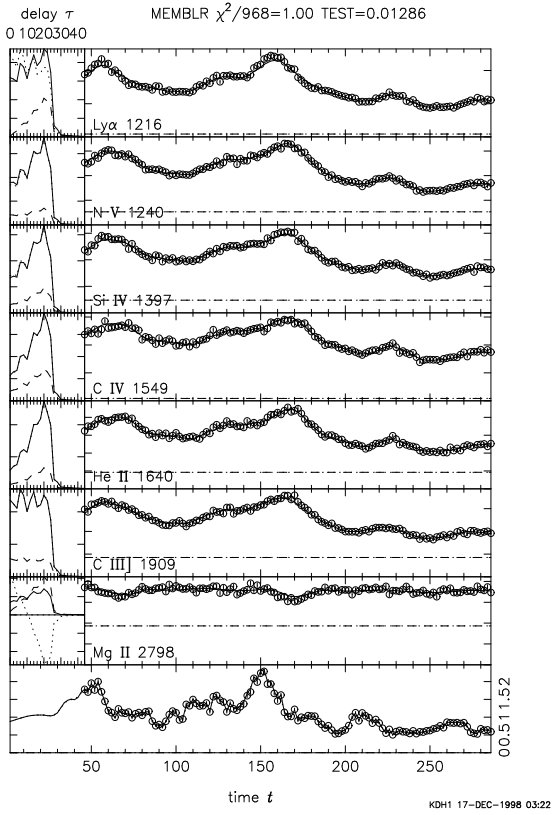}}
\end{picture}\par
\caption[]
{\small
A fit to simulated continuum and emission-line light curves 
for a thin spherical shell.
See text \S\ref{sec:fitslc} for details.
\label{fig:shell60_fit}}
\end{figure}

\subsection{ Fits to Synthetic Light Curves }
\label{sec:fitslc}

Fig.~\ref{fig:shell60_fit} shows our fit 
to the simulated dataset of Fig.~\ref{fig:shell60tru_fit}.
This fit is accomplished,
using the maximum entropy methods discussed in
in Appendix~\ref{app:inverse},
by adjusting the driving light curve $L(t)$,
the background spectrum $F_B(\lambda)$,
and the cloud geometry and density map $f(R,\theta,n_{\sc H})$.
We hold the column density fixed at the correct value
$N_{\sc H}=10^{23}$cm$^{-2}$.
Note that our model does not assume a spherical shell
geometry with all clouds having the same density,
but rather determines what geometry and density distribution
are required in order to fit the light curves.

The predicted light curves are required to achieve a good fit, as
judged by the criterion $\chi^2/N=1$, where $N=968$
is the number of data points.
The fitted model predicts line fluxes that are not quite as high
as the data in the peaks of the light curve, particularly for \lya.
This is the principal defect in the fit. It arises because our
maximum entropy fit seeks the ``smoothest'' functions that
fit the data points.
The reconstructed delay maps, $\Psi_\ell(\tau)$, are noisy at the 10\% 
level, but they do bear a satisfying resemblance to the true delay maps
shown in
Fig.~\ref{fig:shell60tru_fit}.

The maximum entropy fitting techniques
are discussed in detail in  Appendix~\ref{app:inverse}, but
it may be appropriate here to summarise the method and
discuss a few issues.
The fit arises by iterating from initial guesses for the driving
light curve $L(t)$, the background spectrum $F_B(\lambda)$, and the cloud 
map $f(R,\theta,n_{\sc H})$.
The iteration seeks to improve the fit to the data,
aiming for $\chi^2/N=1$,
while also maximizing the ``entropy'' to keep
the model ``as simple as possible'', in a well defined sense.
The iteration either converges or else informs you
that it has failed to converge.
The converged model represents the ``simplest'' model
that fits the data with $\chi^2/N=1$.

The maximum entropy fit is unique in the sense that it does {\em not} 
depend on the initial guesses, and can be reached from a wide 
range of starting points.
However, in a different sense the maximum entropy fit is not
unique because there are a variety of ways to define what you mean by
``as simple as possible''.
For the fits presented in this section, we follow our usual practice
of defining ``simple'' to mean ``smooth''.
Maximizing the entropy ``steers'' each parameter
of the model toward values in adjacent pixels.
This delivers the ``smoothest'' background spectrum, driving 
light curve, and cloud map that succeed in fitting the data.
When two models fit the data equally well, the smoother model is 
preferred.
When the data require a sharp feature in the map,
e.g.\ a peak in the spectrum or light curve, or clouds concentrated at
some radius or with some density, maximizing entropy spreads out
the feature as much as possible in any directions permitted by the data.
In this way the resolution of the map is determined by the
quality of the data, and the sensitivity of the data to each
part of the map.
We will point out examples of this effect below.

\begin{figure}
\vspace*{10.5cm}
\begin{picture}(8,11)
\epsfxsize=8cm
\epsfysize=11cm
 \makebox[8cm][l]{\epsfbox{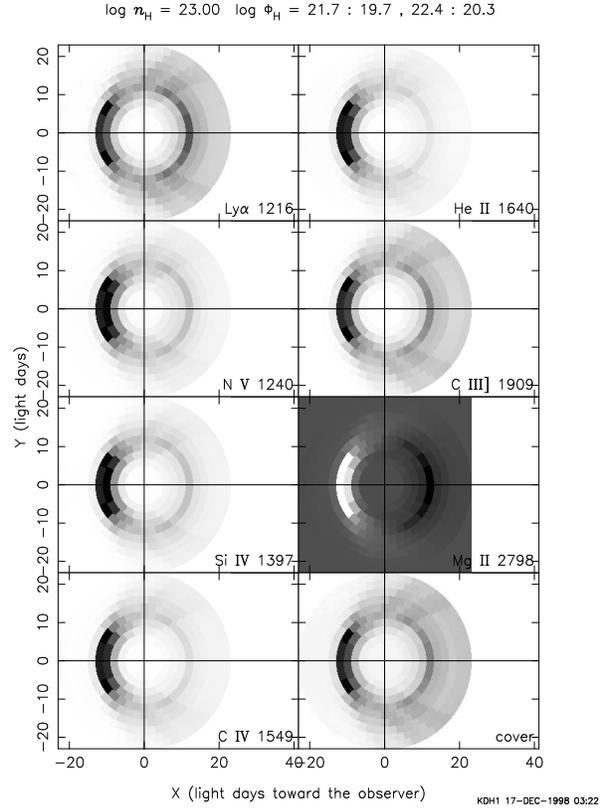}}
\end{picture}\par
\caption[]
{\small
Maps derived from a maximum entropy fit to the simulated data shown 
in Fig.~\ref{fig:shell60_fit}.
The reconstructed geometry $f(R,\theta)$, shown in the lower-right 
panel, reveals the basic form of a hollow spherical shell,
with some smearing along the iso-delay 
parabolas, $\tau \propto R(1+\cos{\theta})$, opening
to the right.
The other panels display maps of the responses
of different emission lines to increases in ionising radiation
from the central source.
The greyscale assigns white and black 
to the minimum and maximum responses occuring in each line.
Positive responses are exhibited by all lines
except \mgii, which has a negative response on the far side 
and a positive response on the near side of the shell.
See text \S\ref{sec:shellxy} for further discussion.
\label{fig:shell60_xy}}
\end{figure}

\begin{figure}
\vspace*{10.5cm}
\begin{picture}(8,11)
\epsfxsize=8cm
\epsfysize=11cm
 \makebox[8cm][l]{\epsfbox{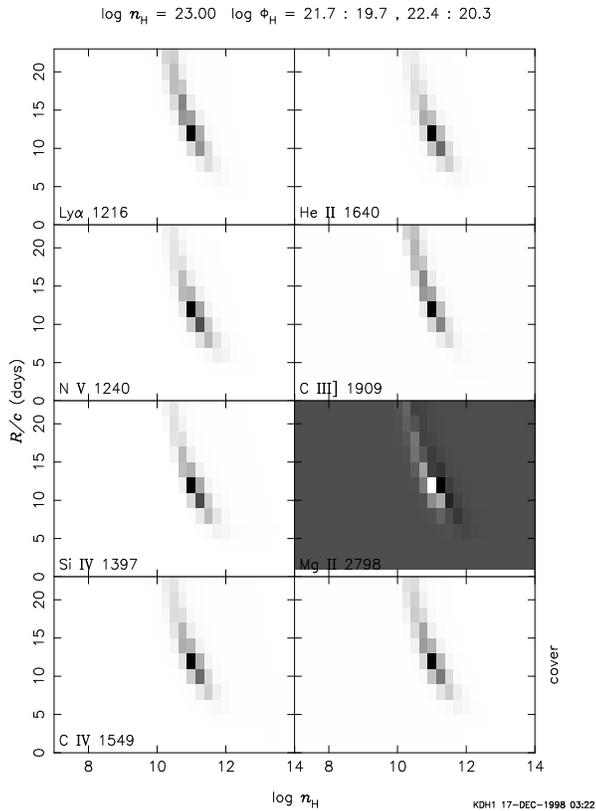}}
\end{picture}\par
\caption[]
{\small
The density-radius map $f(R,n_{\sc H})$
recovered from the fit to the simulated data shown 
in Fig.~\ref{fig:shell60_fit}.
The correct density $n_{\sc H}=10^{11}$cm$^{-3}$ is recovered,
with ambiguity along the direction of constant ionisation
parameter $U \propto R^{-2} n_{\sc H}^{-1}$.

\label{fig:shell60_dr}}
\end{figure}

\subsubsection{ Density--Geometry Maps of the Spherical Shell }
\label{sec:shellxy}

Our maximum entropy fit to the light curves in Fig.~\ref{fig:shell60_fit}
recovers a 3-dimensional map,
$f(R,\theta,n_{\sc H})$, 
specifying the geometry and density distribution of 
the emission-line clouds.
We visualize this by displaying 2-dimensional projections,
$f(R,\theta)$
in Fig.~\ref{fig:shell60_xy} and 
$f(R,n_{\sc H})$
in Fig.~\ref{fig:shell60_dr}, which
we discuss below.

Fig.~\ref{fig:shell60_xy} shows the reconstructed geometry.
The lower-right panel gives the differential covering 
fraction, $f(R,\theta)$, and the other panels show the
responses in the seven ultraviolet emission lines.
The basic morphology of a thin spherical shell,
with radius $R/c \sim 12$ light days,
is clearly recognizable, though distorted in ways that
we understand and discuss below.
The inward anisotropy in most of the lines (stronger response from
the far side of the shell) is recovered.
\mgii\ looks peculiar in the plot because clouds on the far side
of the shell have a negative response in this line. 
A hollow inner region,
$R\approxlt10$ light days, is clearly recovered.
On the far side of the shell,
the cloud distribution is confined to within $\approxlt2$ light
days of the correct radius.
At other azimuths, however, the clouds have spread 
outward along iso-delay parabolas, which open toward the right.
Striations along the iso-delay parabolas are visible
mainly at the top and bottom of the shell.

Fig.~\ref{fig:shell60_dr} shows the density-radius projection,
$f(R, n_{\sc H})$.
Here we see that the correct density, $n \sim 10^{11}$cm$^{-3}$, is 
recovered, though with smearing along the
direction of constant ionisation parameter,
$U \propto R^{-2} n_{\sc H}^{-1}$.
The line ratios evidently constrain $U$ to within about 0.1~dex,
while the constraints on $n_{\sc H}$ and hence $R$ 
are weaker, 0.3 and 0.15~dex respectively.
The data allow clouds can move to a larger (smaller) $R$ provided
they reduce (increase) $n_{\sc H}$ to maintain a fixed 
ionisation parameter, i.e. $n_{\sc H} \propto R^{-2}$.
The line ratios constrain $n_{\sc H}$ because each emission
line becomes optically thin, due to thermalisation or de-excitation,
above a different critical density (e.g. Hamann et al. 2002,
see also Fig.~\ref{fig:shell60_dr}).
This inhibits inward more than outward changes in $R$.

Motion of clouds along iso-delay parabolas
preserves the time-delay constraints imposed by the light curves.
On the near side, clouds constrained by $\tau$ 
can move in $R$ with only a little change in $\theta$,
while on the far side they can move in $\theta$ more than $R$.
This explains why the reconstructed
shell in Fig.~\ref{fig:shell60_xy} is
sharper on the far side than on the near side.
Motion in $\theta$ must also be inhibited to some extent by 
the anisotropic radiation patterns in different lines.

\begin{figure}
\vspace*{10.5cm}
\begin{picture}(8,11)
\epsfxsize=8cm
\epsfysize=11cm
 \makebox[8cm][l]{\epsfbox{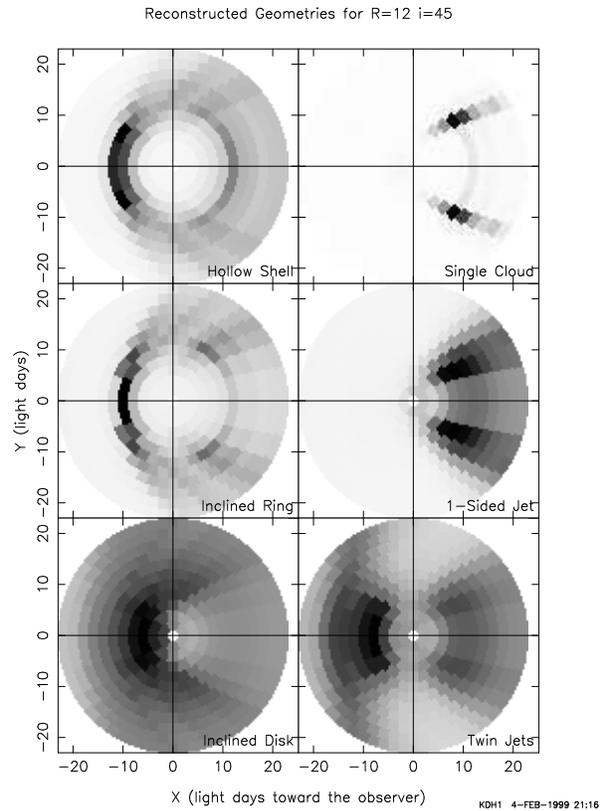}}
\end{picture}\par
\caption[]
{\small
Maps of the six test geometries illustrated
in Fig.~\ref{fig:geometries}
reconstructed from simulated light curves 
with time sampling and signal-to-noise levels
similar to those shown
in Fig.~\ref{fig:shell60_fit}.

\label{fig:rmaps}}
\end{figure}

\subsubsection{ Smooth Maps of Six Geometries }

Encouraged by the results obtained for the spherical shell geometry,
we decided to more thoroughly test the capabilities of quasar tomography
by reconstructing maps from simulated data for a variety of test
geometries.  Fig.~\ref{fig:rmaps} exhibits our reconstructed maps
for the six test geometries illustrated in Fig.~\ref{fig:geometries}.
The maps in Fig.~\ref{fig:rmaps} represent the ``smoothest''
positive maps that fit the data with $\chi^2/N=1$.

Comparison of Figs.~\ref{fig:geometries} and \ref{fig:rmaps} 
indicates that significant ambiguities in the geometry
remain after imposing constraints from the ultraviolet
emission-line light curves.
In all six cases the distorted geometry arises because
the data constrain $U$ and $\tau$ more tightly than 
$n_{\sc H}$, $R$ or $\theta$, as we discussed above
in some detail for the hollow shell geometry.
While the hollow shell geometry is recognizable in the map, 
the map of the inclined ring looks quite similar.
The map of the single cloud is greatly extended along the iso-delay
parabola, making it appear similar to the map of the 1-sided jet.
The 2-sided jet is hardly recognizable, and the disk is totally 
unrecognizable, presumably because these geometries extend over large 
ranges in $R$, $\theta$, and $\tau$. 

The present tests show that the reconstructed geometry
can be significantly distorted while still fitting the lightcurves.
Nevertheless, we consider that even distorted maps such as these
would represent interesting progress given our current limited
understanding of the inner regions of active galactic nuclei.
Furthermore, ambiguities should decrease as more
lines are included in the analysis, since each line responds
with a different sensitivity to $U$ and $n_{\sc H}$.
We also expect to reduce the ambiguity by fitting
velocity-resolved rather than velocity-integrated lightcurves,
since the line ratio constraints are then available separately
for the clouds in each velocity bin, 
rather than only for the sum of all the clouds.
We intend to test these conjectures in future work.
We consider next how to make use of prior information on
the symmetry of the geometry.

\begin{figure}
\vspace*{10.5cm}
\begin{picture}(8,11)
\epsfxsize=8cm
\epsfysize=11cm
 \makebox[8cm][l]{\epsfbox{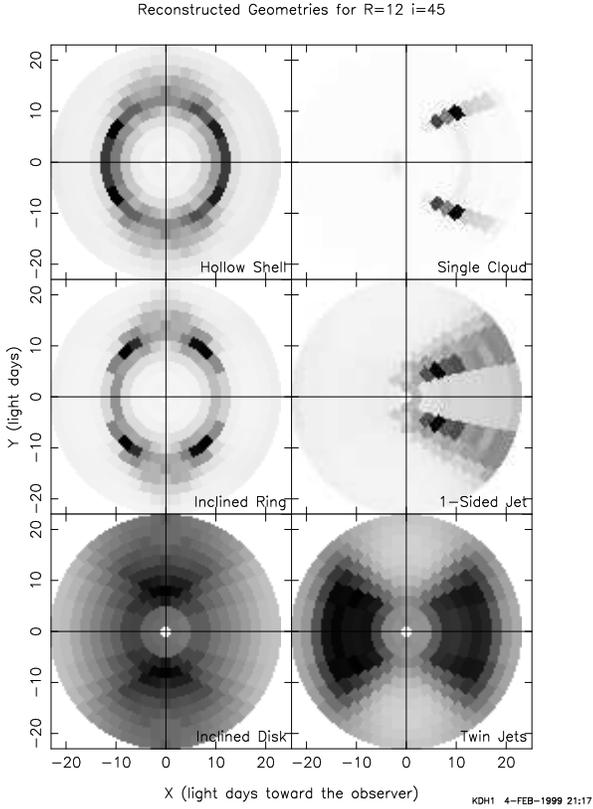}}
\end{picture}\par
\caption[]
{\small
As in Fig.~\ref{fig:rmaps}
except that here the maps are reconstructed with a preference
for front-back symmetry
that is characteristic of point-symmetric and axi-symmetric geometries.
\label{fig:smaps}}
\end{figure}

\subsubsection{ Maps Steered toward Front-Back Symmetry }

The maximum entropy mapping formalism (MEM),
developed in Appendix~\ref{app:inverse},
allows us to incorporate prior information by ``steering''
the fit toward models that obey some type of symmetry.
This is done by defining the entropy in such a way that
a local maximum occurs when the model is symmetric.
MEM then selects the model that fits the data with $\chi^2/N=1$,
and is also ``simple'' in the sense of being ``as close as possible''
to the  desired symmetry.

At present we have relatively little prior information on the
distributions and properties of the emission-line clouds in AGNs,
though this may change in the future.
If we had good reason to assume that the distributions were
spherically symmetric, for example, we would be able to make
better maps by making use of rather than ignoring that prior information.
The maps presented so far are steered toward
blurred copies of themselves, so that the only preferred
symmetry is a local smoothness.
We now consider how to give preference to
geometries with specific symmetries.

It may be helpful to think about this ``steering'' as a way of
asking different questions in order to explore a range
of maps that are consistent with the data.
For example, you may wish to ask ``Are the data
consistent with a spherical geometry?''
MEM delivers an answer to this question in the following way.
If there is a spherically-symmetric map that fits
the data with $\chi^2/N=1$, MEM finds that map.
Usually there will be many spherical geometries,
with different radial profiles, that fit the data.
In that case MEM finds the one that is ``as smooth as possible''
in the radial direction.
If no spherical geometries are consistent with the data,
MEM finds the map that achieves $\chi^2/N=1$ and is 
``as close as possible'' to spherical symmetry.

Steering toward spherical symmetry is very appropriate for the hollow 
shell geometry, and could considerably reduce the distortions
we noted in our reconstructed map.
However, spherical symmetry is not so appropriate for
the other five geometries illustrated in  Fig.~\ref{fig:geometries}.
We therefore opt here for a less restrictive symmetry.
As discussed in \S\ref{sec:geometries},
any point-symmetric or axi-symmetric geometry
has a map that is symmetric in $X$, since for each cloud
on the near side there is a cloud in the corresponding zone
on the far side of the nucleus.
We note that four of the six geometries
in Fig.~\ref{fig:geometries} are symmetric to rotation
about an axis that is inclined to the line of sight.
When these axi-symmetric geometries are rotated around the
line of sight, the result is symmetric between 
the near and far side, $f(X,Y) = f(-X,Y)$.
We may therefore expect to be able to ``improve'' our reconstructions
of these axi-symmetric geometries by ``steering'' them toward
front-back symmetry.
The other two maps, which don't obey this symmetry, will
be altered by this inappropriate steering, but perhaps not by
enough to matter.

Fig~\ref{fig:smaps} exhibits the maps that arise when
preference is given to maps with front-back symmetry.
All six test geometries are now rather more easily
recognizable in the recovered maps.
Smearing is still evident along the parabolic iso-delay
surfaces, but this effect is reduced by the front-back symmetrizing
in comparison with the results in Fig~\ref{fig:rmaps}.
Be reminded that in all cases the maps fit the data with $\chi^2/N=1$.
Differences among the corresponding maps in 
Figs.~\ref{fig:geometries}, ~\ref{fig:rmaps} and \ref{fig:smaps}
illustrate ambiguity in the geometry that is permitted by the data.
The basic resolution is set by the time delay accuracy,
but ambiguity arises primarily because the density is not
well pinned down, allowing the cloud distribution to spread
in radius along iso-delay parabolas while adjusting the density
to hold the ionisation parameter roughly constant.
We suspect that this ambiguity can be reduced
if the dataset includes additional line fluxes that
place tighter constraints on gas densities.

We now briefly discuss each of the maps with a view toward understanding
what might be learned about the geometry by interpreting such maps.
The best reconstruction is that of the hollow shell,
which is quite clearly defined by its map in Fig~\ref{fig:smaps}.
The tight radial resolution ($\approx2$ light days)
that was achieved only
on the back side of the ring ($-X$) in Fig~\ref{fig:rmaps},
is now transferred to the front side ($+X$) as well.
Outward smearing along time-delay parabolas is reduced and
modified by the front-back symmetrizing, leaving the largest
radial smearing ($\approx5$ light days) near the $Y$ axis.
Extrapolating these results, as discussed above, we expect that this 
hollow shell geometry could be very accurately reconstructed if we
steered the model toward spherical symmetry.

The inclined ring should ideally appear as two arcs 
spanning the $Y$ axis (see Fig.~\ref{fig:geometries}).
Its map in  Fig.~\ref{fig:smaps} is superficially
rather similar to that of the hollow shell.
The two geometries could probably be distinguished, however,
based on the azimuthal distribution of the gas around the ring.
In particular, the inclined ring is deficient in clouds near the $X$ axis,
and has four bright points at $(X,Y) = (\pm R\sin{i},\pm R\cos{i})$, 
corresponding to the nearest and
farthest points on the inclined ring.
These features are clearly visible in the map
and would permit an estimate of the ring inclination.
Like the hollow shell map, outward radial smearing is 
expected and evident where the arcs cross the $Y$ axis.

The inclined disk and twin jet maps in  Fig.~\ref{fig:smaps}
are closer to their correct geometries, shown in 
Fig.~\ref{fig:geometries},
but are still not very easily recognisable.
Both maps do indicate cloud distributions spanning a wide range of
radius, in striking contrast to the confined radial range found
in the hollow shell and inclined ring maps.
The inclined disk map has less gas along the $X$ axis
and more along the $Y$ axis, but the
disk geometry and inclination are not clearly identifiable.
If you were to assume a disk geometry, the map implies an
inclination closer to $45^\circ$ than to $0^\circ$
(all clouds on $Y$ axis) or $90^\circ$ (all clouds on $X$ axis).

The twin-jet map has a 
very clear deficit of gas in a wide zone around the $Y$ axis,
with a fairly well defined edge.
However, the two jets blend together across the $X$ axis.
Somewhat better resolution would be required to infer the correct
geometry.

The single cloud and 1-sided jet geometries
have clouds exclusively on the near side of the nucleus,
with no corresponding clouds on the far side.
This violates the preferred front-back symmetry employed
in reconstructing the maps in Fig~\ref{fig:smaps}.
Maximizing the entropy with the preferred symmetry
encourages these maps to develop spurious far-side copies
of their near-side features.
However, it seems that even in these highly asymmetric cases
the maps are not much affected,
and are quite similar to or perhaps somewhat better
than those in Fig~\ref{fig:rmaps}.
The quality of the light curve data constraints must be sufficient
in this example to rule out responses at larger time delays,
thus over-riding the preferred symmetry.

\section{ Quasar Distances }
\label{sec:cosmology}

So far we have assumed that the distance of the active galaxy is known.
If the distance is incorrect, the assumed luminosity will be
incorrect and the resulting maps will be distorted.
How sensitive are the maps to the adopted distance?
To put this question the other way, can we use our maps
to determine distances and luminosities
of active galaxies?   If so, then emission-line reverberation
studies could provide a new method to measure cosmological
parameters such as $H_0$ and $q_0$.

The observed flux from an active nucleus at redshift $z$ is
\begin{equation}
F(\lambda) \equiv \lambda F_\lambda(\lambda) 
= \frac{ L(\lambda_e) }{ 4 \pi D_L^2}
= \frac{ h c }{ \lambda_{\sc H} }
	\frac{ Q_{\sc H} S(\lambda_e) \zeta(i) }{ 4 \pi D_L^2 }
\ ,
\end{equation}
where $L(\lambda) \equiv \lambda L_\lambda(\lambda)$
is the luminosity,
$\lambda_e = \lambda/(1 + z)$ is the emitted wavelength,
\begin{equation}
D_L = \frac{cz}{H_0} \left[
	1 + \frac{ (1 - q_0) z }{ 1 + q_0 z + \sqrt{ 1 + 2 q_0 z } } \right]\ ,
\end{equation}
is the luminosity distance,
where $H_0$ is the Hubble constant
and $q_0$ is the deceleration parameter,
$Q_{\sc H}$ is the emission rate of hydrogen-ionising photons,
\begin{equation}
	S(\lambda)
	= \frac{ \lambda_{\sc H} }{h c} \frac{ L(\lambda) }{ Q_{\sc H} }
	= \frac{ \lambda_{\sc H} L(\lambda) }
{ \int_0^{\lambda_{\sc H}} L(\lambda) d\lambda } 
\end{equation}
is the dimensionless shape of the photon spectrum,
and $\zeta(i)$ is a dimensionless continuum anisotropy factor,
allowing the spectrum to depend on the observer's viewing angle $i$.

As we have seen, by using
photoionisation models to fit
measured emission-line flux ratios we
can in principle constrain 
the hydrogen density $n_{\sc H}$ and
the ionisation parameter $U = \Phi_{\sc H}/(n_{\sc H}c)$,
where $\Phi_{\sc H}$ is the hydrogen-ionising flux 
incident on the emission-line gas.
Since $\Phi_{\sc H} = Q_{\sc H} / 4\pi R^2 $,
we can estimate the ``photoionisation radius''
\begin{equation}
R_Q = \left( \frac{ Q_{\sc H} }
	{4 \pi \Phi_{\sc H}} \right)^{1/2}
= D_L \left( \frac{ \lambda_{\sc H} F(\lambda) }
	{ h c \Phi_{\sc H} S(\lambda) \zeta(i) } \right)^{1/2}
\ .
\end{equation}
Echo mapping experiments give an independent measurement
of the radius,  ``reverberation radius'',
\begin{equation}
	R_\tau = \frac{c\ \tau}
	{ (1+z)(1 + \cos{\theta}) }
\ ,
\end{equation} 
which is derived from time delay measurements.
Equating $R_Q$ and $R_\tau$ then gives
the luminosity distance as
\begin{equation}
 D_L = \left( \frac{ h c \Phi_{\sc H} S(\lambda_e) \zeta(i) }
	{ \lambda_{\sc H} F(\lambda) } \right)^{1/2}
	\frac{ c \tau }
	{ ( 1 + z )( 1 + \cos{\theta}) }\ .
\end{equation} 
The Hubble constant (for small $z$) is then
\begin{equation}
H_0 \approx \frac{cz}{D_L}
\approx \left( \frac{ \lambda_{\sc H} F(\lambda) }
	{ h c \Phi_{\sc H} S(\lambda_e) \zeta(i) } \right)^{1/2}
\left( \frac{ z ( 1 + \cos{\theta} ) } 
{\tau } 
\right) 
\ .
\end{equation}

\subsection{ Distance--Geometry Correlations }

Most of the quantities on the right-hand side of the above expressions
for $D_L$ and $H_0$ are constrained by observations:
$\lambda F_\lambda$ and $z$ are directly observed, 
$\Phi_{\sc H}$ is determined
from the emission-line flux ratios, $\tau$ is determined
from the light curve time delays.
The spectral shape $S(\lambda)$ may be estimated from the spectral energy
distribution obtained from multi-wavelength observations extending
from X-rays to the infrared.  The extreme ultraviolet range is not directly
observable, and this introduces some uncertainty.  Extinction and
reddening by dust in the host galaxy and in our own Milky~Way
must also be taken into account, of course.

The possibility of anisotropic ionising radiation,
$\zeta(i) \neq 1$,may be more difficult to diagnose.
Unification scenarios suggest that we view Seyfert 1 galaxies
near the pole, $i\approxlt60^\circ$, and Seyfert 2 galaxies
at higher inclinations so that a thick dusty torus obscures our
view of the broad emission line region.
Large equivalent widths of emission lines would imply that the
photoionised clouds see a stronger continuum than we do.
This effect may be useful in estimating $\zeta$, which
appears to be similar in different objects,
judging from the similarity of their emission-line equivalent widths.

Ambiguity arises also from $\cos{\theta}$.  We expect 
$\left<~\cos{\theta}~\right>=0$ for any 
point- or axi-symmetric geometry.
However, gas could be located preferentially on the far
side or near side of the nucleus.  Very good light curves may constrain
the shape of the time delay map $\Psi(\tau)$, rather than just the
mean value of $\tau$, and this would constrain $\cos{\theta}$.  The
value of $\cos{\theta}$ is also constrained by the different
anisotropies exhibited by different emission lines.  Thus there would
appear to be some prospect for constraining $\cos{\theta}$, although
we do not expect this constraint to be very strong unless the data are
very good.

Considering the uncertainty in $\cos{\theta}$, we expect an ambiguity
to arise between $H_0$ and the inferred geometry.  A large value of
$H_0$ (or equivalently a lower source luminosity) may be accommodated
by decreasing $R$ and increasing $\cos{\theta}$, i.e. moving gas
inward and placing it on the far side of the nucleus.  Because
$1+\cos{\theta}$ can be at most 2, this degree of freedom should allow
up to a factor of 2 increase in the inferred value of $H_0$. In practice this
can only be realised for an extreme geometry with all gas located on
the far side of the nucleus. Conversely, a
small value of $H_0$ can be accommodated by increasing $R$ and
decreasing $\cos{\theta}$, thus moving gas toward the observer and out
to larger radii so that once again, $\tau$ is roughly preserved.

\begin{figure}
\vspace*{10.5cm}
\begin{picture}(8,11)
\epsfxsize=8cm
\epsfysize=11cm
 \makebox[8cm][l]{\epsfbox{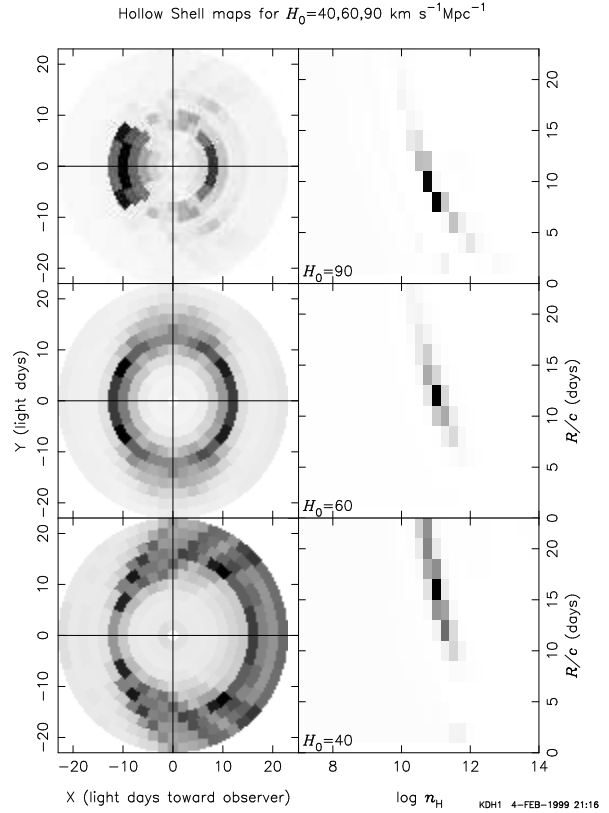}}
\end{picture}\par
\caption[]
{\small
Maps of a spherical shell with $R/c=12$d and $n=10^{11}$cm$^{-3}$
recovered from fits to the simulated data shown 
in Fig.~\ref{fig:shell60_fit}.
The Hubble constant used to generate the fake data
was $H_0=60$~km~s$^{-1}$Mpc$^{-1}$,
while the three reconstructions assume
$H_0=40,60,90$~km~s$^{-1}$Mpc$^{-1}$.
\label{fig:shell3hs}}
\end{figure}

\subsection{ $H_0$ from Simulated Data }

To investigate prospects for determining $H_0$,
and to verify the expected
correlations outlined above, we performed
tests with simulated light curves.
We generated fake light curves assuming
$H_0=60$~km~s$^{-1}$Mpc$^{-1}$, and then
reconstructed the gas distribution using various
values of $H_0$ to see how that affected the fit to the data
and the resulting cloud maps.
When we assumed a value of $H_0$ fairly close to the correct value,
it was possible to achieve a good fit to the light curve,
as judged by $\chi^2/N=1$, but with a distorted geometry.
When we assumed a very discrepant value of $H_0$,
it became impossible to achieve $\chi^2/N=1$.
By using the entropy to quantify the distorted geometry,
we could identify the correct value of $H_0$ as that
giving the best fit with the simplest geometry.

Representative results are presented in Fig.~\ref{fig:shell3hs}
for a geometrically thin spherical shell geometry,
with $R/c=12$~light days, hydrogen density $n_{\sc H}=10^{11}$~cm$^{-3}$,
and hydrogen column density $N_{\sc H}=10^{23}$~cm$^{-2}$.
The geometry and density maps shown
for $H_0=40,60,90$~km~s$^{-1}$~Mpc$^{-1}$
are reconstructed by fitting to light curves generated
for $H_0=60$~km~s$^{-1}$~Mpc$^{-1}$.
In all three cases the fit achieves $\chi^2/N=1$.
The thin-shell geometry is fairly accurately recovered
when we assume $H_0=60$~km~s$^{-1}$~Mpc$^{-1}$.
For $H_0=90$~km~s$^{-1}$~Mpc$^{-1}$,
the map is highly distorted with most of the gas from the shell
displaced to smaller $R$ and away from the observer to form a clump
on the far side of the nucleus.
For $H_0=40$~km~s$^{-1}$~Mpc$^{-1}$, the map is distorted in
the opposite sense, with gas displaced away from the nucleus
and toward the observer along iso-delay parabolas.

The correct value, $H_0=60$~km~s$^{-1}$~Mpc$^{-1}$,
may be recognized as the value giving the most symmetric map.
The uncertainty in $H_0$ resulting from such fits may be
quantified in terms of the posterior probability distribution
derived from the fit
\begin{equation}
	P(H_0|\vec{D}) \propto P(\vec{D}|H_0)\ P(H_0) .
\end{equation}
Here $P(H_0)$ is the prior probability assigned to different
values of $H_0$ before we consider the data $\vec{D}$.
Regardless of our prior view on the value of $H_0$,
the evidence provided by the data $\vec{D}$
multiplies our prior probability by the factor
\begin{equation}
	P(\vec{D}|H_0) \propto {\rm exp}\left\{ \alpha S - \chi^2/2 \right\}\ .
\end{equation}
This applies for each value of $H_0$ the appropriate penalties
for failing to fit the data (large $\chi^2$)
and for deploying excessive numbers of parameters
(large negative $ \alpha S$, where $S < 0$ is the entropy of the map
and $\alpha > 0$ is the mixing parameter).
For more detailed discussion, see Appendix~\ref{app:inverse}.

\section{ Summary and Concluding Remarks }
\label{sec:summary}

We have developed a method of reconstructing 5-dimensional
maps of photoionised emission-line regions by fitting predictions
of photoionisation models to high-quality observations of
reverberating emission-line spectra of active galactic nuclei.
Reverberation effects in the emission lines offer
information about the $(R,\theta)$ geometry and 
the angular emission pattern of the gas clouds,
coded as time-delay information in the light curves of different lines
with respect to the continuum.
At the same time,
line profiles give the distribution
of clouds with line-of-sight velocity $v$,
line ratios constrain the density $n_{\sc H}$
and column density $N_{\sc H}$ of the gas clouds,
and equivalent widths constrain the covering fraction $f$.
Our method brings all these constraints together
in a global fit to reverberating emission-line spectra
in order to recover a 5-dimensional cloud map,
the differential covering factor
$f(R,\theta,n_{\sc H},N_{\sc H}, v)$.
This method may allow high-quality observations 
to reveal the geometry, physical conditions, and kinematics of the 
emission-line gas.

We use the maximum entropy method to find
the ``simplest'' positive maps (maximum entropy)
that fit the data ($\chi^2/N=1$).
The entropy is measured with respect to a default map, 
giving us some freedom to choose what we consider to be a 
``simple'' map.
We use the entropy to give preference to smooth maps 
rather than noisy ones, and to ``steer'' the map toward various
possible symmetries -- spherical, axial, point, etc.
The range of maps found for different choices of symmetry
helps to gauge the extent and nature of ambiguities 
that remain after fitting the data.
The Bayesian formulation provides posterior 
probability distributions, and Monte Carlo techniques
may also be used to assess statistical uncertainties
in the cloud map, the distance, and any other parameters 
that affect the fit.

Tests with simulated datasets indicate good prospects for mapping geometry
and density structure from observations of emission-line light curves
that show evidence for time delays.  
Good results require simultaneous fits to numerous lines
in order to constrain both the ionisation parameter and the density.
The main ambiguity arises because the emission-line lightcurves
place tighter constraints on the ionisation prameter $U$
and time delay $\tau$ than on the geometry $(R,\theta)$
and physical conditions $(N_{\sc H},n_{\sc H})$.
Our simulation tests suggest that in some cases the correct geometry
is recognizable in the reconstructed maps
when we give preference to front-back symmetry,
a property of both point-symmetric and axi-symmetric cloud distributions.
If geometries are indeed approximately symmetric, then
distances can be derived in order to
constrain cosmological parameters, ($H_0$ and $q_0$).

The test cases presented in this paper consider only a single
column density, $N_{\sc H} = 10^{23}$cm$^{-2}$.
A range of $N_{\sc H}$ at each radius will probably be
required to fit observed lightcurves
(Shields, Ferland, Peterson 1995).
Extending the photoionisation grid to include a range of $N_{\sc H}$ 
is straightforward, but will require more computer time
for the iterative fitting.
The additional freedom should also increase the ambiguity,
but perhaps not too severely because
in general the emission line spectra of clouds
in the range $10^{22} < N_{\sc H} < 10^{24}$ 
are less sensitive to $N_{\sc H}$ than to 
$U$ and $n_{\sc H}$
(e.g. Korista et~al. 1997; Goad \& Koratkar 1998).

We have largely omitted to discuss mapping the kinematics of
the emission-line gas, focussing instead on mapping the
geometry by fitting to velocity-integrated emission-line light curves.
Extracting kinematic information will involve simultaneous
fitting of light curves at many wavelengths to resolve time variations
and hence time delays at each velocity in the emission-line profile.
Extending our analysis to include multiple pixels along the $v$--axis
of the cloud map is straightforward.
Note that continuum fitting and de-blending of lines in the observed spectra
will not be necessary because the model
adds together the predicted continuum and the velocity profiles
of all the lines.
It will be interesting to see the extent to which $v-R$ projections of 
$f(R,\theta,n_{\sc H},N_{\sc H}, v)$ provide evidence for virial 
motions.
We expect the use of velocity information to reduce the ambiguities,
since constraints will then be available separately for many 
subsets of the cloud population rather than only for the
sum over all clouds.

We rely on CLOUDY both to generate our fake datasets and to
reconstruct from them the cloud maps.
In the analysis of real observations,
uncertainties in the atomic data, in the elemental abundances
in the shape of the ionising continuum, 
and in the assumptions and approximations
used in the photoionisation calculations
will each give rise to errors and distortions of the maps.
It will therefore be important to continue to improve the
atomic data and photoionisation physics 
in tandem with attempts to fit CLOUDY predictions to observations.

Finally, we note that the methods developed here for
interpreting spectral variations in active galactic nuclei
can also be extended to other types of astronomical objects
in which an unknown gas distribution is excited by photoionisation
from a compact source of ionising radiation.
In cases where time-dependent reverberation effects are not observed,
the observed spectrum alone or, better still,
spatially-resolved spectral information from 
combinations of narrow-band imaging and long-slit or integral-field
spectroscopy can be used as constraints.
Maximum entropy fitting then recovers
the ``simplest'' viable maps of the geometry, kinematics, and physical
conditions in the photoionised gas.
A likely future application is simultaneous fitting of
narrow-band images and long-slit spectra of nova shells
and planetary nebulae.

\section*{acknowledgments}

We thank Brad Peterson and Gary Ferland for ideas, 
inspiration, and helpful comments on the manuscript.
PPARC grants supported MRG and KTK at St.Andrews
during the period of most rapid progress on this paper.
KH acknowledges support from a PPARC Senior Fellowship.

\appendix

\section{ The Forward Problem }
\label{app:forward}

\subsection{ Geometry }

Two parameters, $\Delta t$ and $\tau_{\sc max}$,
set the pixel size and radius of the region that we wish to map.
These need to be chosen appropriately for each dataset.
We recommend setting $\Delta t$ to the median time step between 
successive data points, and $\tau_{\sc max}$ to a value
larger than the largest delay for which the data provide evidence
of significant response, usually not more than a third
of the full duration of the observed light curves. 
The model includes a constant background spectrum, \S{\ref{sec:bgspec}},
to allow for responses on larger timescales.
Our maximum entropy fits (Appendix~\ref{app:inverse})
degrade the resolution of the map as much as possible
within the constraints set by the light curve data.
If the pixels are too large, the resolution may be limited
by the pixel size rather than by the data.
If the pixels are too small, the resolution is defined by
the data, but more computer time is required
to accomplish the fit.
Thus high or low signal-to-noise datasets may require
smaller or larger $\Delta t$, respectively.

\subsubsection{ Shells }

In the model developed for this paper, we divide the spatial volume
surrounding the nucleus into $N_R$ concentric spherical shells with
radii
\begin{equation}
	R(i) = i \Delta R
\ , 
\end{equation}
where $\Delta R = c \Delta t / (1+z)$
and $\Delta t$ is a suitable time interval
comparable to the time spacing of the observations.
The time delay range covered by the shells is thus
$\tau_{\rm min} = 0$ to $\tau_{\rm max} = 2 N_R \Delta t$.
Any flux arising from outside this region is included in the
model in the form of a constant background spectrum
(\S\ref{sec:spectra}),
thus neglecting any reverberation effects.
Note that other partitions are possible, for example equal
spacing in $\log{R}$ may be more appropriate if a
large range of radius is to be mapped.

The exact choice of $\Delta t$ is not too critical because
in fitting the data we first construct predicted light curves
uniformly sampled with a time interval $\Delta t$,
and then interpolate to the actual times of observations.
Given an equally spaced time series,
$\Delta t$ can be the time interval
between successive data points.
For unequally-sampled time series,
$\Delta t$ can be the minimum or median time
spacing, or some point in between.
A smaller $\Delta t$ may permit resolution of finer
structure if the data record significant flux changes
from one data point to the next.
A larger $\Delta t$ may be adequate if the data record
no significant flux changes between successive data points.

\subsubsection{ Zones }

The time delay accounting for light travel time is
\begin{equation}
 \tau(R,\theta) = \frac{R}{c} (1+z) \left( 1 + \cos{\theta} \right)\ ,
\end{equation}
where the angle $\theta$ specifies the direction of the gas
cloud as seen from the nucleus measured from the direction
away from the observer.
Note that the time delay depends on $\theta$ as well as $R$,
as do the reprocessing efficiencies discussed
in \S\ref{sec:dew} below.
We therefore sub-divide shell $i$ into $N_\theta(i) = 2i+1$
equal-area zones delimited by equal intervals of $\cos{\theta}$.
The solid angle covered by each zone as viewed from the origin is
\begin{equation}
	\Delta\Omega = \frac{ 4 \pi } {N_\theta}
	= \frac{ 4 \pi }{ 2i + 1 }\ .
\end{equation}
The cosine of our viewing angle for clouds in zone $j$ of shell $i$ is then
\begin{equation}
	\cos{\theta} = 2 \left( \frac{j-1}{N_\theta-1} \right) - 1
	= \frac{ j - i - 1 }{i}\ .
\end{equation}
Finally, the time delay is
\begin{equation}
	\tau = 2 i \Delta t \left( \frac{j-1}{N_\theta-1} \right)
	= \Delta t \left( j - 1 \right)\ .
\end{equation}
Note that because we choose $N_\theta = 2i+1$,
the delay spacing is $\Delta t$ in all shells.
The full range of delays is $0 < \tau < \tau_{\sc max} = 2 N_R \Delta t$.

Because none of the observable data we consider
\footnote{
Polarimetry data might provide a means of resolving structure in $\phi$.
} 
depend upon the angle $\phi$ measured
around the line of sight, we can omit further sub-division of the
zones into sectors.

We wish to emphasize that our use of a shell and zone partition of
the 3-dimensional volume is completely general and does not
exclude any class of models.  While we employ nested shells,
this does not imply a restriction to a spherically-symmetric geometry,
since structure in the angle $\theta$ is resolvable by the zones that
partition each shell.  We do not partition the zones in the angle
$\phi$, but this does not restrict us to models that are symmetric to
rotation around the line of sight.  It is just that we are unable to
detect any $\phi$ structure that may exist because there are no
aspects of the data that depend upon the angle $\phi$.  What we can
detect, then, is the true 3-dimensional geometry $f(R,\theta,\phi)$
projected by rotation around the line of sight to give a 
2-dimensional map
\begin{equation}
	f(R,\theta) = \int f(R,\theta,\phi) d\phi\ .
\end{equation}

\subsubsection{ Cloud covering fraction }

Zone $\theta$ of shell $R$ subtends solid angle $\Delta\Omega$ and contains
a population of gas clouds that intercepts a fraction 
$f(R,\theta)\ \Delta R\ \Delta \Omega / 4\pi$
of the ionising luminosity.  
The cloud cover
arises from different types of clouds characterized by their hydrogen
density $n_{\sc H}$, column density $N_{\sc H}$, and line-of-sight
velocity component (Doppler shift) $v$.  
The cloud geometry is
therefore obtained by integrating over the cloud types :
\begin{equation}
f(R,\theta) = \int f(R,\theta,n_{\sc H},N_{\sc H},v)\
	dn_{\sc H}\ dN_{\sc H}\ dv\ .
\end{equation}
This introduces the 5-dimensional differential covering fraction
map that we aim to reconstruct by
fitting to the observed spectral variations.
Since $f$ is a distribution, we can use any convenient partition
to divide the $(R, \theta, n_{\sc H}, N_{\sc H}, v )$  domain.
In this paper we employ a partition with equal intervals of
$R$, $\cos{\theta}$, $\log{n_{\sc H}}$, $\log{N_{\sc H}}$, and $v$.

The radial dimension of a cloud of density $n_{\sc H}$ and column $N_{\sc H}$
is $l_R \sim N_{\sc H}/n_{\sc H}$.
To fit within its spherical shell, the cloud
must satisfy $N_{\sc H}/n_{\sc H} < \Delta R$.
For $\Delta R = 10^{15.4}$cm (1 light day),
this requires $N_{\sc H} < 10^{26.4}$cm$^{-2} (n_{\sc H}/10^{11}$cm$^{-3})$.
In this way we may justify excluding clouds from a corner
of the $n_{\sc H}-N_{\sc H}$ plane, although our present
implementation does not impose this constraint.

If $y>0$ is a cloud's tangential to radial aspect ratio,
we may consider clouds that are spherical ($y=1$),
radially-elongated ``cigars'' ($y<1$), 
or radially-flattened ``pancakes'' ($y>1$).
The tangential area covered by such a cloud 
is  $A \sim y^2 (N_{\sc H}/n_{\sc H})^2$, and
the differential covering fraction for the ensemble of such clouds
is then
\begin{equation}
f(R,\theta,n_{\sc H},N_{\sc H},v) \sim \int
	y^2 \left( \frac{N_{\sc H}}{n_{\sc H}}\right )^2
		\frac{ \Delta R\ \Delta \Omega }{ 4\pi }\ n_{\rm cl}\ dy
\ ,
\end{equation}
where 
$n_{\rm cl}(R,\theta,n_{\sc H},N_{\sc H},v,y)$
is the volume number density of clouds.
Thus our cloud map $f(R,\theta,n_{\sc H},N_{\sc H},v)$
does not assume spherical clouds but rather represents the
combined sky coverage of whatever cloud shapes
are present within the volume.
Note, however, that in
\S\ref{sec:aim} we adopt an anisotropy function that may be more 
appropriate for spherical clouds than for cigars or pancakes.

\subsubsection{ Cloud shadowing }

Our photoionisation code CLOUDY assumes 
an unobstructed path from the origin out to each cloud. 
If a cloud is present at position $(R,\theta,\phi)$,
self-consistency demands that clouds at the same $\theta,\phi$
but larger $R$ should be unable to ``see'' the centre.
\footnote{
The ionising spectrum incident upon the clouds is clearly a combination of 
the ``bare'' spectrum from the nucleus together with
transmitted, reflected, and diffuse spectral components
arising from other clouds.
Since fully transparent clouds do not contribute to the overall
covering fraction calcuation, 
covering fractions of greater than 1 are possible.
}
For our shell-zone geometry, this constraint can be satisfied if
\begin{equation}
	f(\theta) = \int_0^{R_{\sc MAX}} f(R,\theta)\
	\frac{\Delta\Omega\ \Delta R}{4\pi}
	\leq 1\ .
\end{equation}
At present we do not implement this as a hard constraint upon the solution,
but rather we use the entropy to ``steer'' the map toward this constraint
(see Appendix~\ref{app:inverse}),
and check the final result.

\subsection{ Ionising Radiation }

\begin{figure}
\begin{picture}(8,7)
\put(0,0){\includegraphics{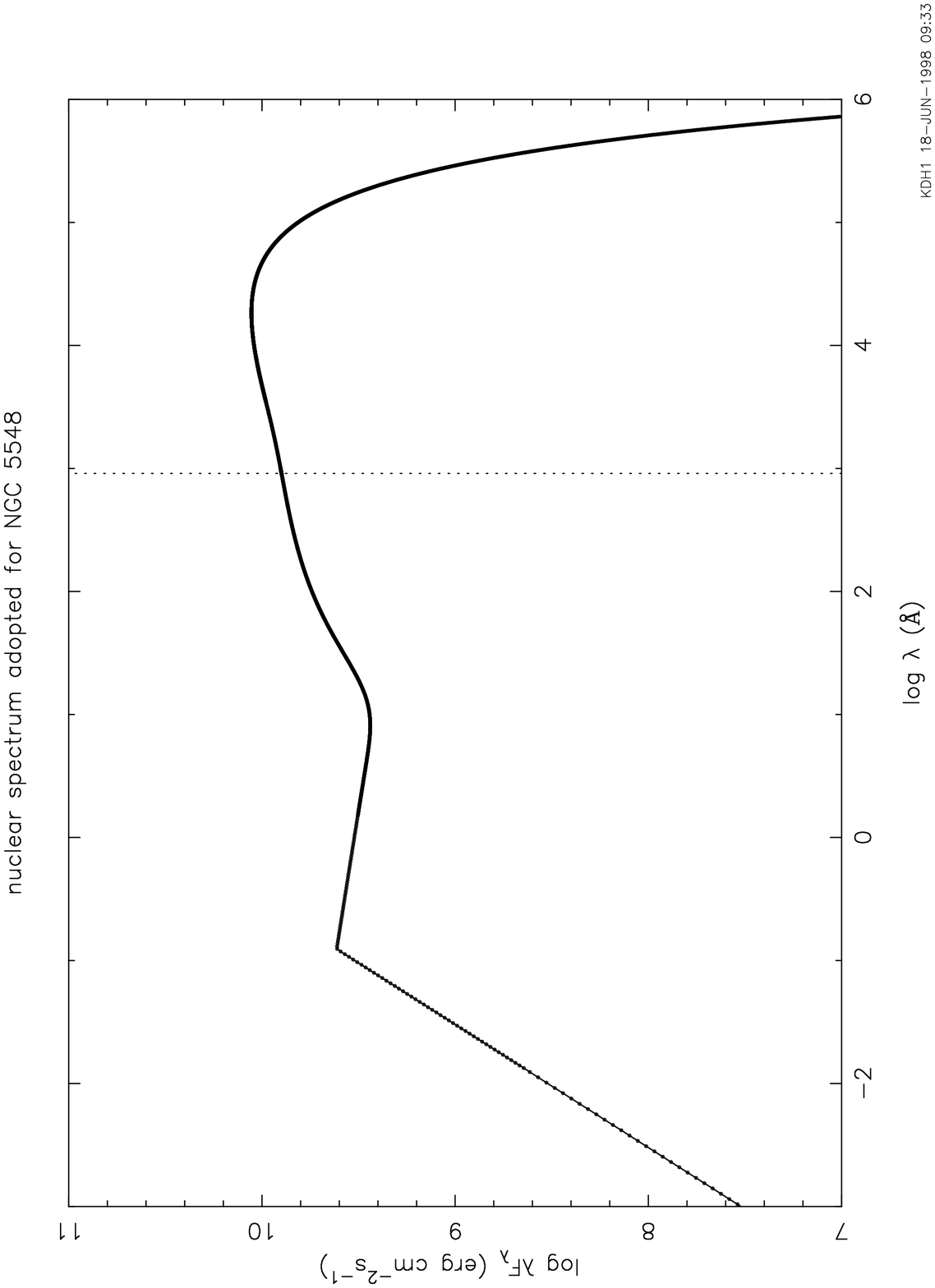} }
\noindent
\end{picture}
\vspace*{7cm}
\par
\caption[]
{\small
The nuclear spectrum adopted in photoionisation modelling of NGC~5548
and in the simulation tests presented in this paper.
\label{fig:inci_5548}}
\end{figure}

We assume that the nucleus is a point source emitting
a time-variable spectrum described by
\begin{equation}
L(\lambda,t) \equiv \lambda L_\lambda(\lambda,t) = 
L_0 S(\lambda) L(t)\ .
\end{equation}
Here the time variations of the nucleus are specified by
the dimensionless light curve $L(t)$,
the shape of the nuclear spectrum is given by
the dimensionless spectrum
\begin{equation}
S(\lambda) = \frac{ \lambda_{\sc H} L(\lambda,t) }
{ \int_0^{\lambda_{\sc H}} L(\lambda,t) d\lambda }
\ ,
\end{equation}
and the dimensional luminosity scale factor is
$L_0 = Q_H h c / \lambda_H$,
where $Q_{\sc H}$ is the rate at which
the nucleus emits hydrogen-ionising photons
($\lambda < \lambda_{\sc H} = 912$\AA)
in the reference state $L(t)=1$.
This separable form, with $S(\lambda)$ independent of $t$,
assumes spectral shape does not change
as the luminosity varies in time.
The assumption can obviously be modified if required.
The specific shape adopted for the
calculations in this paper is shown in Fig~\ref{fig:inci_5548}.

The flux of hydrogen-ionising photons incident on a gas
cloud located a distance $R$ from the nucleus is
\begin{equation}
\Phi_{\sc H}(t) = 
\frac{ Q_{\sc H} L(t-\tau) }{ 4\pi R^2 }\ .
\end{equation}
Here $t$ is the time at which we receive the reprocessed
photons from the gas cloud,
and the time delay $\tau$ accounts for the longer light travel time
on the path from the nucleus to the gas cloud to the observer
compared with the direct path from the nucleus to the observer.

\subsection{ Time-Dependent Spectra  }
\label{sec:spectra}

The spectrum that we observe at wavelength $\lambda$ and time $t$ is
modelled as the sum of three components,
a time-independent background spectrum, $F_B(\lambda)$,
time-variable direct light from the nucleus, $F_D(\lambda,t)$
and reprocessed light,  $F_R(\lambda,t)$,
arising from gas clouds near the nucleus
that are responding to ionising radiation from the nucleus:
\begin{equation}
F(\lambda,t) \equiv
\lambda F_\lambda(\lambda,t) 
	= F_B(\lambda)
	+ F_D(\lambda,t)
	+ F_R(\lambda,t) 
\ .
\end{equation}

\subsubsection{ Background Spectrum }
\label{sec:bgspec}

The background spectrum  $F_B(\lambda)$ is included in the
model to account for sources of light that
are effectively constant on the timescale spanned by the
observed light curves.
Such sources include starlight from the host galaxy or nuclear 
starburst, and reprocessed light from gas outside the
domain of the map that respond to the ionising radiation
but on timescales much longer than the duration of the light curves.

\subsubsection{ Direct Light }

The direct light is
\begin{equation}
	F_D(\lambda,t) = F_0\ S(\lambda_e)\ L(t)
\ ,
\end{equation}
where 
$S(\lambda)$ and $L(t)$ are the dimensionless
spectral shape and time variations,
and
$F_0 = L_0 / 4 \pi D_L^2$ is the dimensional flux
observed in the reference state with $L(t)=1$.
Note that with the source at redshift $z$
we must evaluate the spectral shape at the emitted wavelength
$\lambda_e = \lambda (1+z)$, and that
$D_L$ is the luminosity distance, e.g.
\begin{equation}
D_L = \frac{cz}{H_0} \left[
	1 + \frac{ (1 - q_0) z }{ 1 + q_0 z + \sqrt{ 1 + 2 q_0 z } } \right]\ ,
\end{equation}
where $H_0$ is the Hubble constant
and $q_0$ is the deceleration parameter.

\subsubsection{ Reprocessed Light }

The reprocessed light that we see at time $t$, arises from
gas at $(R,\theta)$ responding to the ionising flux from the
nucleus at the earlier time $t-\tau$, where 
\begin{equation}
\tau = \frac{R}{c} (1+z) \left( 1 + \cos{\theta} \right)\ .
\end{equation}
The gas cloud therefore sees an incident hydrogen-ionising photon flux
\begin{equation}
\Phi_{\sc H} = \frac{ Q_{\sc H} L(t-\tau) }{ 4\pi R^2 }\ .
\end{equation}

We model the spectrum of the reprocessed light as
\begin{equation}
F_R(\lambda,t) =
	F_0 S(\lambda_0) 
	\int_0^\infty L(t-\tau) 
	\Psi( \tau, \lambda | L(t-\tau) )
	d\tau\ .
\end{equation}
Here $\lambda_0=1215$\AA\ is a reference wavelength,
discussed in \S{\ref{sec:dew}} below.
Note that this model accounts for non-linear reprocessing
by allowing the delay map $\Psi(\tau,\lambda | L)$
to depend on $L$,
the dimensionless ionising luminosity of the nucleus.
Non-linear responses arise because the reprocessing efficiencies
of the gas clouds change when they are exposed to different levels
of ionising flux.
The delay map
is obtained by summing contributions from all types of clouds,
each contributing at the appropriate delay $\tau$,
and weighted by the differential covering fraction $f$
and reprocessing efficiency $\epsilon$:
\begin{equation}
\begin{array}{l}
\Psi( \tau, \lambda | L ) = 
 \int \frac{ d\Omega }{4\pi} dR\ dn_{\sc H}\ dN_{\sc H}\ dv\
\\ ~~~~ \times
  f(R,\theta,n_{\sc H},N_{\sc H},v)\
\\ ~~~~	\times
  \epsilon( \lambda, \Phi_{\sc H}, n_{\sc H}, N_{\sc H}, v, \theta )\
\\ ~~~~  \times
\delta\left\{ \tau - \frac{R}{c}(1+z)(1+\cos{\theta}) \right\}\
.
\end{array}
\end{equation}
Here 
$\epsilon( \lambda, \Phi_{\sc H}, n_{\sc H}, N_{\sc H}, v, \theta )$
is the dimensionless reprocessing efficiency, discussed below,
and the Dirac $\delta$ function ensures that the correct 
time delay is used at each reprocessing site.

\subsection{ Reprocessing Efficiencies }
\label{sec:dew}

The reprocessing efficiencies, $\epsilon_c$ for the continuum 
and $\epsilon_\ell$ for
line $\ell$, are defined in terms of dimensionless equivalent widths,
i.e. the reprocessed flux emerging from the cloud divided by the
incident flux $\lambda F_\lambda$ at the reference wavelength 
$\lambda_0 = 1215$\AA.  
These we evaluate with the photoionisation code
CLOUDY (Ferland et al. 1998)
for a pre-specified grid in $\log{\Phi_{\sc H}}$, 
$\log{n_{\sc H}}$, and $\log{N_{\sc H}}$,
for given abundances and shape of ionising spectrum
(Korista et al. 1997).
Linear interpolation in the grid is then used to evaluate
the efficiencies as required.

The reprocessed light includes both line and continuum emission:
\begin{equation}
\begin{array}{c}
\epsilon( \lambda, \Phi_{\sc H}, n_{\sc H}, N_{\sc H}, v, \theta ) =
	( \lambda_0 / \lambda ) 
	\epsilon_c( \lambda, \Phi_{\sc H}, n_{\sc H}, N_{\sc H}, \theta )
\\	+ \sum_\ell \lambda_0 G_\lambda( x_\ell ) 
	\epsilon_\ell( \Phi_{\sc H}, n_{\sc H}, N_{\sc H}, \theta )\ .
\end{array}
\end{equation}
For the reprocessed continuum light, 
$\epsilon_c( \lambda, \Phi_{\sc H}, n_{\sc H}, N_{\sc H},
\theta)$ gives the ratio of $\lambda F_\lambda$ for the reprocessed
continuum flux divided by that for the incident flux at $\lambda_0$.
For the emission lines 
$\epsilon_\ell( \Phi_{\sc H}, n_{\sc H}, N_{\sc H}, \theta)$ 
gives the flux in
line $\ell$ emerging in direction $\theta$ divided by the incident flux
$\lambda_0 F_\lambda(\lambda_0)$.

When spectra are required, the line emission is assigned a Gaussian
distribution in wavelength
\begin{equation}
G_\lambda(x_\ell) = \frac{ 
{\rm exp}\left\{ -\frac{1}{2}x_\ell^2 \right\} 
}{ \sqrt{2\pi} \Delta\lambda }\ ,
\end{equation}
where
\begin{equation} 
x_\ell = \frac{ (1+z) \lambda 
	- (1+v/c) \lambda_\ell }{ \Delta\lambda } 
\end{equation} 
allows for the Doppler shift by velocity $v$ from the rest wavelength
$\lambda_\ell$.  The line width $\Delta \lambda$ represents the spectral
resolution of the data and intrinsic broadening, e.g., Doppler
broadening due to thermal velocities and the range of velocities
$\Delta v$ in the velocity bin.

\begin{figure}
\vspace*{10.5cm}
\begin{picture}(8,11)
\epsfxsize=8cm
\epsfysize=11cm
 \makebox[8cm][l]{\epsfbox{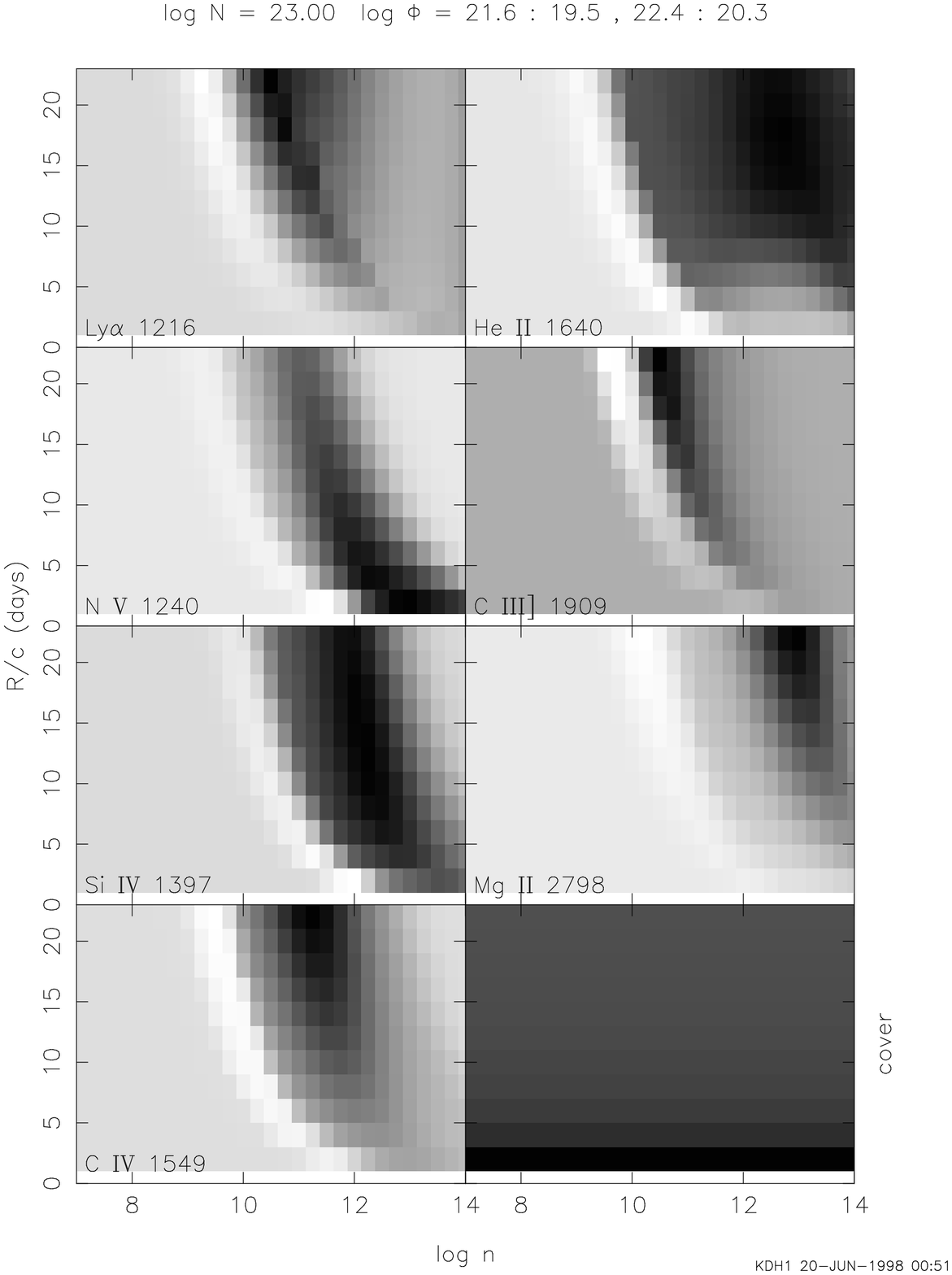}}
\end{picture}\par
\caption[]
{\small
The response as a function of density and radius is shown for each of
the major ultraviolet emission lines for a filled spherical volume
that reprocesses equal fractions of the ionising spectrum in each unit
of radius.
\label{fig:sphere60_dr}}
\end{figure}

Fig.~\ref{fig:sphere60_dr} examines the responses of the major
ultraviolet emission lines to changes in the flux of ionising photons.
The figure shows for each radius and density the difference in the
line flux between bright and faint states that correspond to an
increase by 0.8 in $\log{\Phi_{\sc H}}$.  In this LOC model, we
consider a single column density $\log{N_{\sc H}}=23$~cm$^{-2}$,
and a range
of densities $ 7 < \log{n_{\sc H}} < 14$~cm$^{-3}$.  The radial range
considered is $2 < R < 24$ light days, corresponding roughly to a
range in the ionising photon flux of $22 > \log{\Phi_{\sc H}} > 20$.

Each emission line reprocesses efficiently in a somewhat different
region of of the $n_{\sc H}$-$R$ plane.  The regions are delimited to first
order by an appropriate range of the ionisation parameter $U \propto
\Phi_{\sc H}/n_{\sc H} \propto Q_{\sc H}/(R^{2}n_{\sc H})$.  
For constant $n_{\sc H}$,
the reprocessing efficiency diminishes at large $\Phi_{\sc H}$ (small $R$)
because the clouds become fully ionised and heat to Compton
temperatures, and at small $\Phi_{\sc H}$ (large $R$) because only a
surface layer on the inward face of the cloud is ionised.  Efficient
reprocessing requires a higher density at small $R$, i.e. $n_{\sc H}
\propto Q_{\sc H}/(R^2U)$.

In the LOC model, a range of $n_{\sc H}$ is present at each $R$, and so the
reprocessing regions can simply move outward as the ionising photon
flux increases.  Clouds on the inner edge of the reprocessing region
become less efficient while those on the outer edge become more
efficient reprocessors.  Lining the inner edge of the reprocessing
region is a region of negative response, where increasing the ionising
photon flux decreases the reprocessed line flux (see Goad et~al. 1993).

At constant $U$, reprocessing efficiencies decrease at large $n_{\rm
H}$ where collisional processes compete with radiative de-excitation
of the upper levels of the transitions.  This trend is seen in all
lines except \nv, whose response increases with increasing $n_{\rm
H}$.
The \nv\ emission line, for solar abundances and the 
spectral energy distribution assumed here,
begins thermalising above $\sim10^{13}$~cm$^{-3}$.
If we extended our grid to smaller radius and higher density,
the \nv\ emission would decrease again,
appearing as \civ\ does above $\sim10^{11}$~cm$^{-3}$.

\subsubsection{ Anisotropic Emission }
\label{sec:aim}

The reprocessed radiation is generally emitted with an anisotropic
angular distribution, typically because emission is produced on
the inward face of an optically thick cloud
(Ferland et al. 1992).
To allow for anisotropic
radiation, the inward efficiency $\epsilon_I$ and outward efficiency $\epsilon_O$
are computed separately.  For the angular pattern of the reprocessed
light we assume
\begin{equation}
\epsilon(\theta) = \bar{\epsilon}
	+ \left( \epsilon_I - \bar{\epsilon} \right) \cos{\theta}\ ,
\end{equation}
where the mean efficiency is
\begin{equation}
	\bar{\epsilon} = ( \epsilon_I + \epsilon_O ) / 2\ .
\end{equation}
The linear dependence on $\cos{\theta}$ corresponds to isotropic
radiation from each point on the surface of a spherical cloud, 
with the inward and outward fluxes assigned to the inward and outward
facing hemispheres.

\begin{figure}
\vspace*{10.5cm}
\begin{picture}(8,11)
\epsfxsize=8cm
\epsfysize=11cm
 \makebox[8cm][l]{\epsfbox{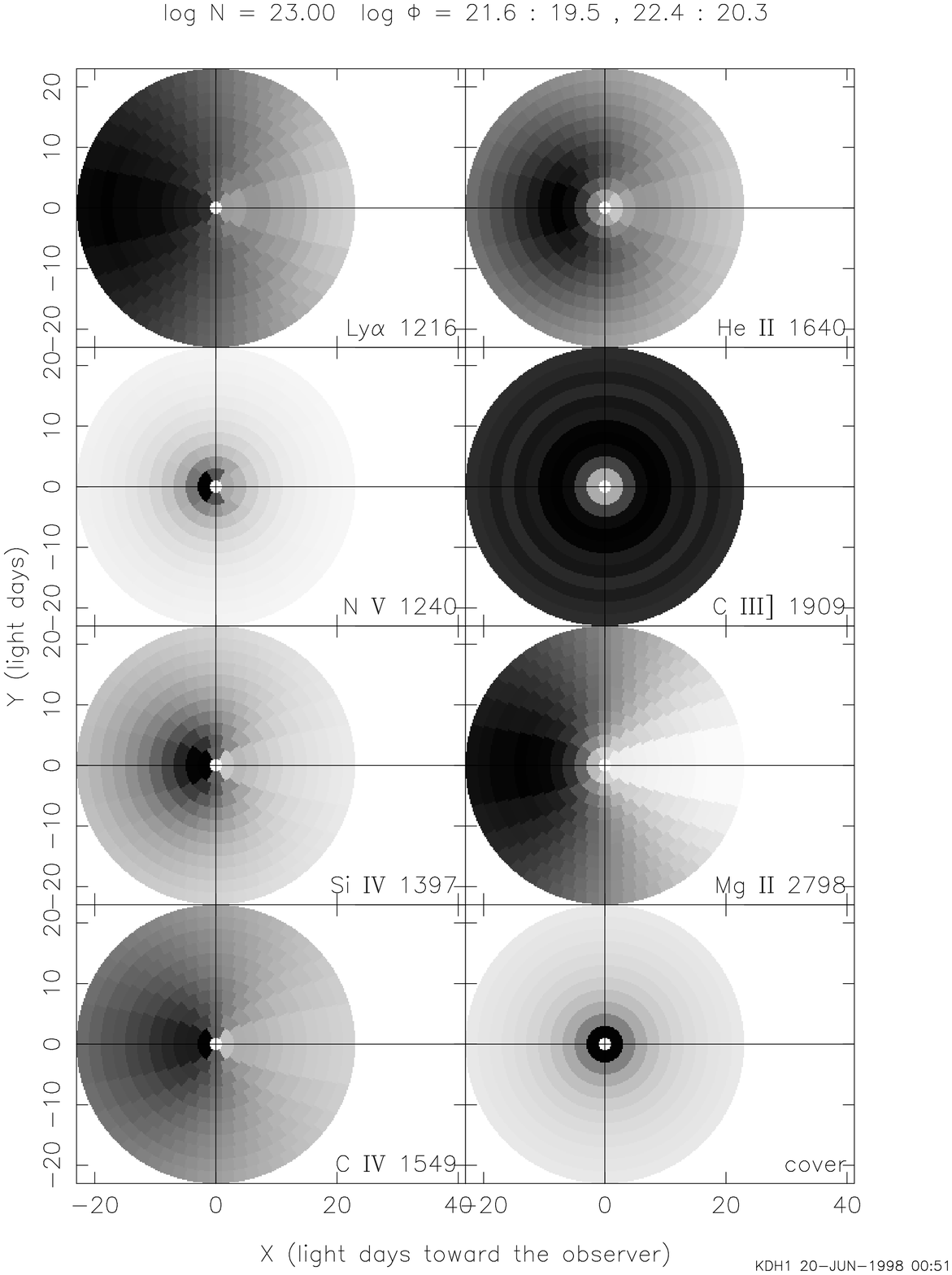}}
\end{picture}\par
\caption[]
{\small
Maps showing the appearance in each of the major ultraviolet
emission lines for a filled spherical volume that reprocesses
equal fractions of the ionising spectrum in each unit of radius.
The inward anisotropy in certain lines is evident.
\label{fig:sphere60_xy}}
\end{figure}

Fig.~\ref{fig:sphere60_xy} shows maps for each of the major
ultraviolet emission lines for a filled spherical volume that
reprocesses the same fraction of the ionising radiation in each shell.
Note that there is a sum over density to obtain the total 
response in each pixel of the map. 
The sum over density
includes clouds with both positive and negative response.  The clouds
with positive responses evidently dominate, so that in no part of the
emission-line region is there a net negative response.

An inward anisotropy (stronger response from the far side
of the map) is evident for most of the lines.
This arises from high-density clouds that are optically thick.
The exception is \ciii], which we treated as isotropic
in the present calculation.
We expect the \ciii] inward/total ratio to be 0.5-0.6 for clouds that are 
emissive in this line (O'Brien, Goad \& Gondhalekar~1994),
and we plan to include this small anisotropy in future calculations.

\subsection{ Delay Maps }

\begin{figure}
\vspace*{10.5cm}
\begin{picture}(8,11)
\epsfxsize=8cm
\epsfysize=11cm
 \makebox[8cm][l]{\epsfbox{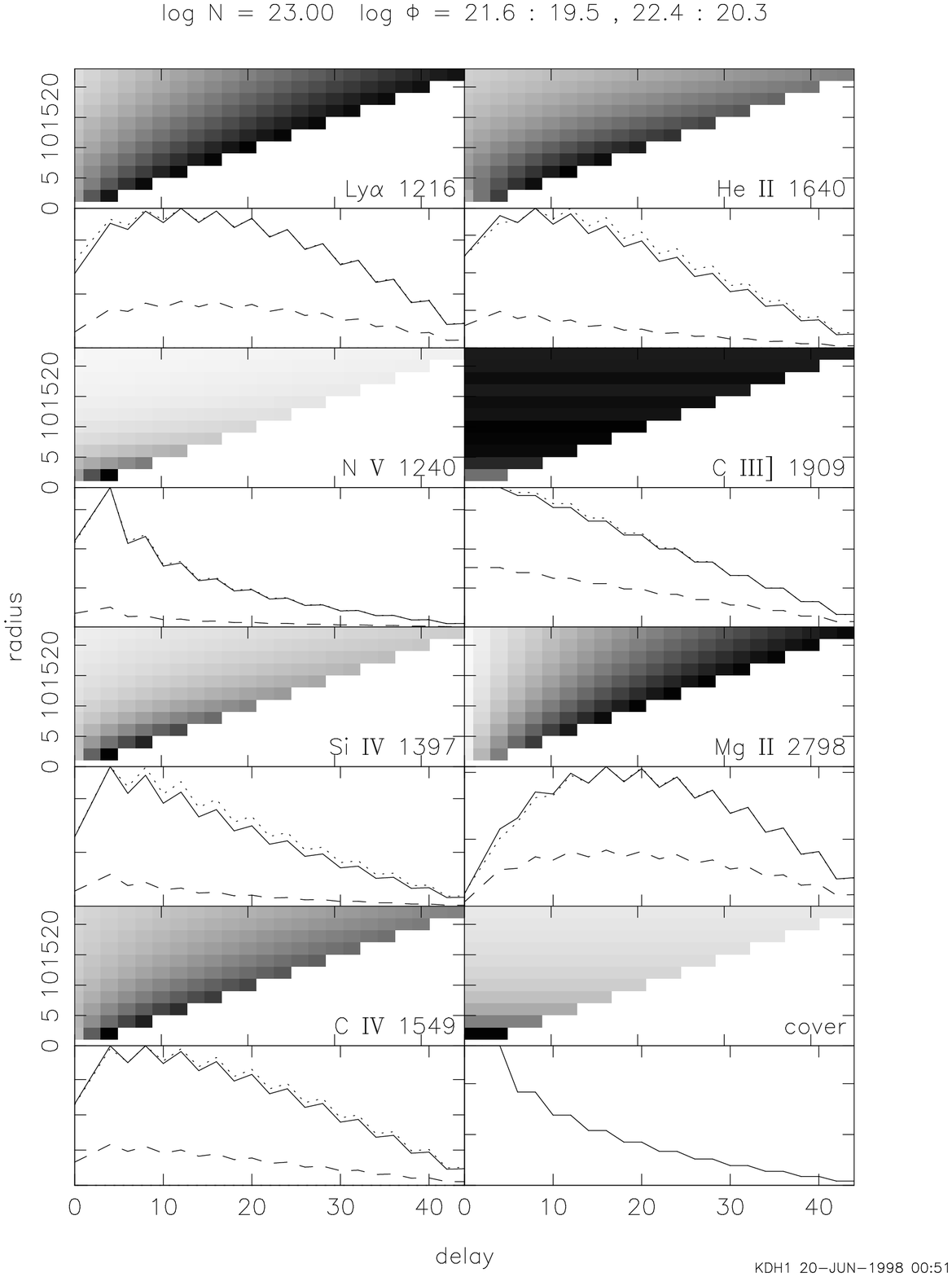}}
\end{picture}
\par
\caption[]
{\small
Delay maps for each of the major ultraviolet emission lines
for a filled spherical volume that reprocesses
equal fractions of the ionising spectrum in each unit of radius.
Each shell produces a wedge of emission extending over
$0 < \tau < 2R/c$ with a slope that increases with the inward anisotropy.
\label{fig:sphere60_tr} }
\end{figure}

Delay maps $\Psi(\tau)$ are shown in Fig.~\ref{fig:sphere60_tr}.  Note
that each spherical shell makes a wedge shaped contribution to
$\Psi(\tau)$ in the region $0 < \tau < 2R/c$.  The near side of each
shell contributes at $\tau=0$ while the far side contributes at
$\tau=2R/c$.  The slope of the wedge reflects the degree of anisotropy
of the line emission, with steeper slopes indicating increased
inward anisotropy.
The filled sphere
is thus a sum of wedge-shaped contributions from the various shells.

For each emission line the response has been evaluated in both a
bright and a faint state, and the corresponding delay maps are shown
as solid and dashed curves.  Also shown as a dotted curve is the
difference between the bright and faint state delay maps, scaled to
the same peak value.  The dotted curve would be identical to the solid
curve if the response is strictly linear, and this is evidently
closely satisfied for all of the lines.

The linearity of the responses and the lack of negative responses
seen in this LOC model are consistent with the two implicit assumptions
normally made when maximum entropy techniques are used to recover delay
maps from observed light curves
(e.g., Horne, Welsh \& Peterson 1991).

\subsection{ Synthetic Light Curves }

\begin{figure}
\vspace*{10.5cm}
\begin{picture}(8,11)
\epsfxsize=8cm
\epsfysize=11cm
 \makebox[8cm][l]{\epsfbox{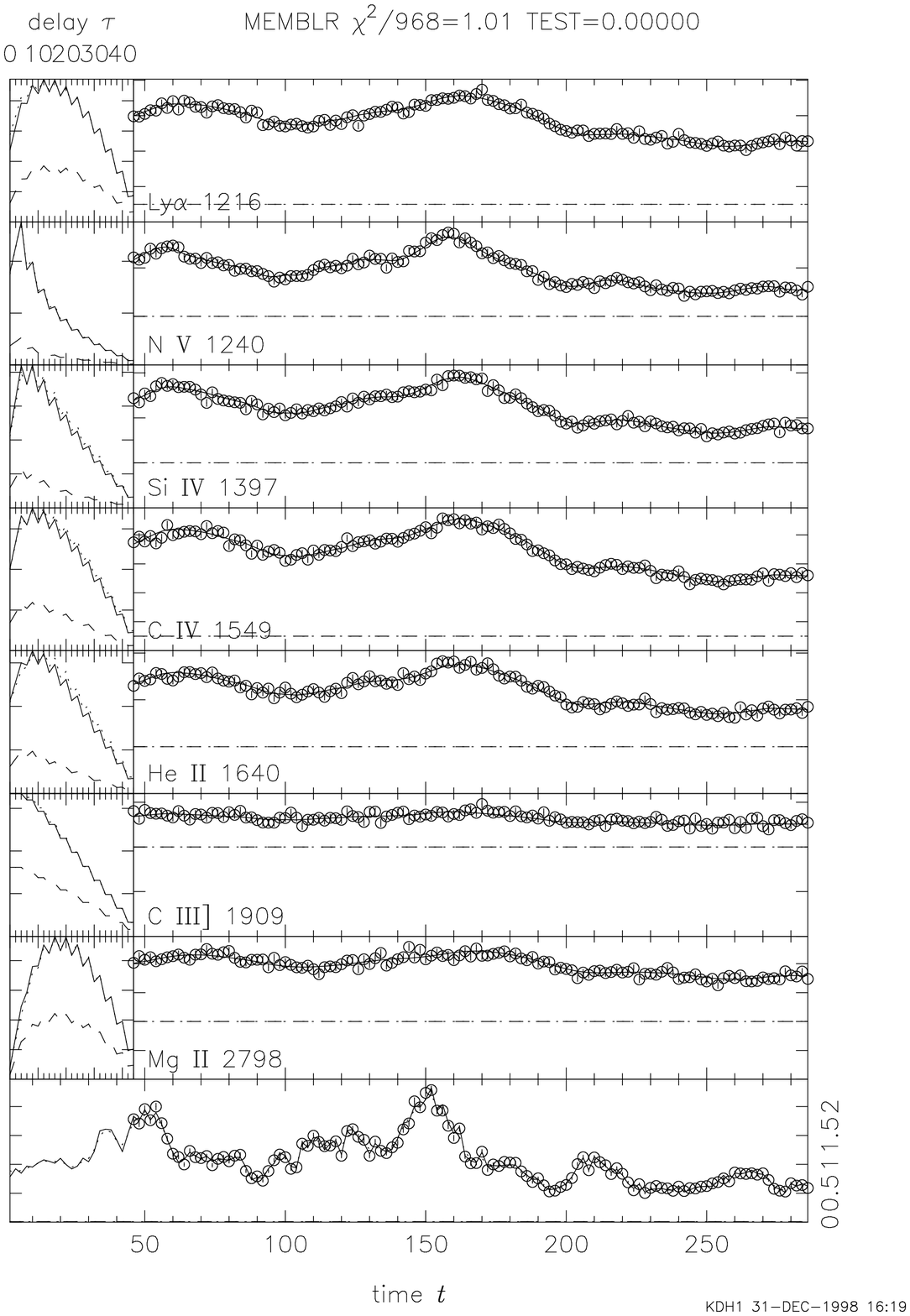}}
\end{picture}
\par
\caption[]
{\small
Synthetic light curves for the LOC model.
\label{fig:start_fit} }
\end{figure}

Fig.~\ref{fig:start_fit} shows synthetic light curves calculated
for the LOC model.
The bottom panel shows the driving light curve, which was generated
assuming that log of the flux executes a random walk in time.
The emission-line light curves in the upper panels exhibit
variations correlated with those in the driving light curve,
but delayed and smeared due to the appropriate delay maps
shown to the left of each emission-line light curve.

\subsection{ Synthetic Spectra }

\begin{figure}
\begin{picture}(8,7)
\put(0,0){\includegraphics{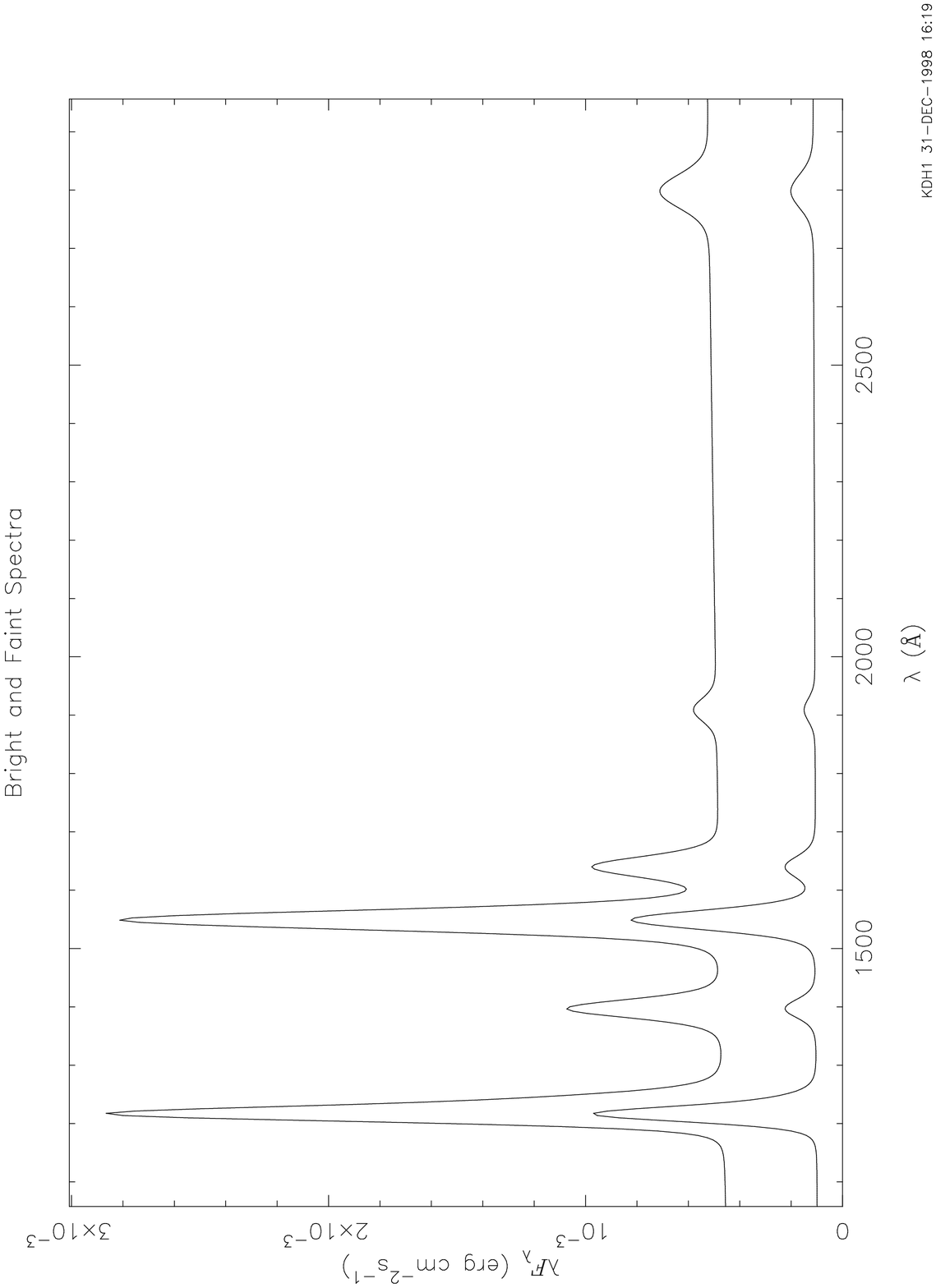} }
\noindent
\end{picture}
\vspace*{7cm}
\par
\caption[]
{\small
Synthetic bright and faint spectra for the LOC model.
\label{fig:locspec} }
\end{figure}

Fig.~\ref{fig:locspec} shows synthetic spectra calculated for the LOC
model in the bright and faint states.  
The velocity
distribution at each radius is taken to be a Gaussian with dispersion
$\Delta v = \sqrt{GM/R}$ for $M=3 \times 10^7M_\odot$.
The model spectra bear a satisfying resemblance
to the observed spectra of Seyfert~1 galaxies,
for example NGC~5548 (Clavel et al.~1991).
In particular, the \civ/\lya\ ratio is close to 1.
The main differences are significantly 
weaker \mgii\ and \ciii] emission in the predicted spectrum.
From Fig.~\ref{fig:sphere60_xy}, we see that the  \mgii, \ciii],
and \lya\ emission regions extend to the 20 light day outer radius
assumed in this LOC model.
These lines would be stronger if we increased
the outer radius.

\section{ The Inverse Problem }
\label{app:inverse}

Appendix~\ref{app:forward} 
dealt with the relatively straightforward problem of
calculating predicted data $F_\lambda(\lambda,t)$ given a particular
cloud map $f(R,\theta,n_{\sc H},N_{\sc H},v)$.
We now tackle the more treacherous
inverse of this problem, fitting an observed dataset to recover
the underlying cloud map.  We accomplish this using maximum
entropy fitting methods similar to those we have developed
for other astro-tomography problems,
for example eclipse mapping (Horne~1985),
Doppler tomography (Marsh \& Horne 1988),
and echo mapping (Horne, Welsh \& Peterson~1991, Horne 1994).

Our discussion of the inverse problem is rather formal because 
the maximum entropy fitting techniques are based on
probability theory and therefore very generally applicable
rather than being ``ad-hoc'' methods limited to our specific problem.
Specific examples of maximum entropy fitting applied to the
quasar tomography problem are discussed in
\S\ref{sec:cloudmaps}.

We adopt the notation $\vec{f}$ to refer to a vector (or map) of
parameters $f_i$, with $i$ labeling the axes of the parameter space
(or the pixels of the map).  
In quasar tomography, $\vec{f}$
includes several positive additive distributions:
the cloud map $f(R,\theta,n_{\sc H},N_{\sc H},v)$,
the light curve $L(t)$,
the background spectrum $F_B(\lambda)$,
and the distance $D$.
In practice of course we adopt a partition for the domain of
each distribution, thereby replacing the infinite-dimensional
distribution by a finite-dimensional map $\vec{f}$.

Constraints on $\vec{f}$ arise from measurements of a data vector
$\vec{D}$, which impose a probability distribution $P(\vec{D})$.
In quasar tomography, $P(\vec{D})$ arises from the incomplete and noisy
measurements of the time-variable spectrum $F_\lambda( \lambda, t )$.
The forward problem, calculating predicted data $\vec{R}(\vec{f})$
given the map $\vec{f}$, is well defined (Appendix~\ref{app:forward}).
The inverse problem,
estimating $P(\vec{f})$ given $P(\vec{D})$, is often
under-constrained because the number of parameters $M={\rm dim}(\vec{f})$ 
exceeds the number of data constraints $N={\rm dim}(\vec{D})$.
This indeterminacy can be particularly severe when high-quality datasets
warrant the use of fine partitions to resolve small structures in the
maps.

To cope with the indeterminacy of the inverse problem, we regularize
the problem by seeking ``the simplest'' or ``most probable''
solutions that fit the observations.
This is a good example of ``Occam's razor'':
when two models succeed equally well in accounting for the data,
the simpler model is more likely to be true.

\subsection{ To Fit the Data: Maximize Likelihood }

Bayes's theorem tells us how to assess the probability of
a model with parameters $\vec{f}$ in the light of a dataset $\vec{D}$.
The {\it posterior probability} of $\vec{f}$ is
\begin{equation}
	P(\vec{f}|\vec{D}) =
	\frac{ P(\vec{D}|\vec{f})\ P(\vec{f}) } {P(\vec{D})}\ .
\end{equation}
Here the {\it likelihood}, $P(\vec{D}|\vec{f})$,
is the probability that $\vec{D}$ arises
if we assume that $\vec{f}$ is true.
The {\it prior}, $P(\vec{f})$, is the probability
assigned to $\vec{f}$ before $\vec{D}$ was obtained.
Finally, the divisor
\begin{equation}
P(\vec{D}) = \int P(\vec{D}|\vec{f})\ P(\vec{f})\ d\vec{f}
\end{equation}
is included to ensure that $P(\vec{f}|\vec{D})$
is a probability density normalized to 1 
when integrated over all possible maps $\vec{f}$.

The likelihood is easy to evaluate.
To be specific about the dataset $\vec{D}$,
consider $N$ independent measurements $D_i$,
described by Gaussians with standard deviations $\sigma_i$
centred on predicted values $R_i(\vec{f})$.
The likelihood is then
\begin{equation}
P(\vec{D}|\vec{f}) = 
\prod_{i=1}^N P(D_i|\vec{f}) = 
	\frac{ {\rm exp}\left\{-\chi^2/2 \right\} } { Z_D }\ ,
\end{equation}
where
\begin{equation}
\chi^2 = \sum_{i=1}^N \left( \frac{D_i - R_i(\vec{f}) }{\sigma_i}\right)^2\ ,
\end{equation}
and the partition function,
\begin{equation}
Z_D = \int {\rm exp}\left\{-\chi^2/2 \right\} d\vec{D}
	= \prod_{i=1}^{N} \left( 2 \pi \sigma_i^2 \right)^{1/2}\ ,
\end{equation}
normalizes the likelihood as a probability density on $\vec{D}$.

\begin{figure}
\vspace*{10.5cm}
\begin{picture}(8,11)
\epsfxsize=8cm
\epsfysize=11cm
 \makebox[8cm][l]{\epsfbox{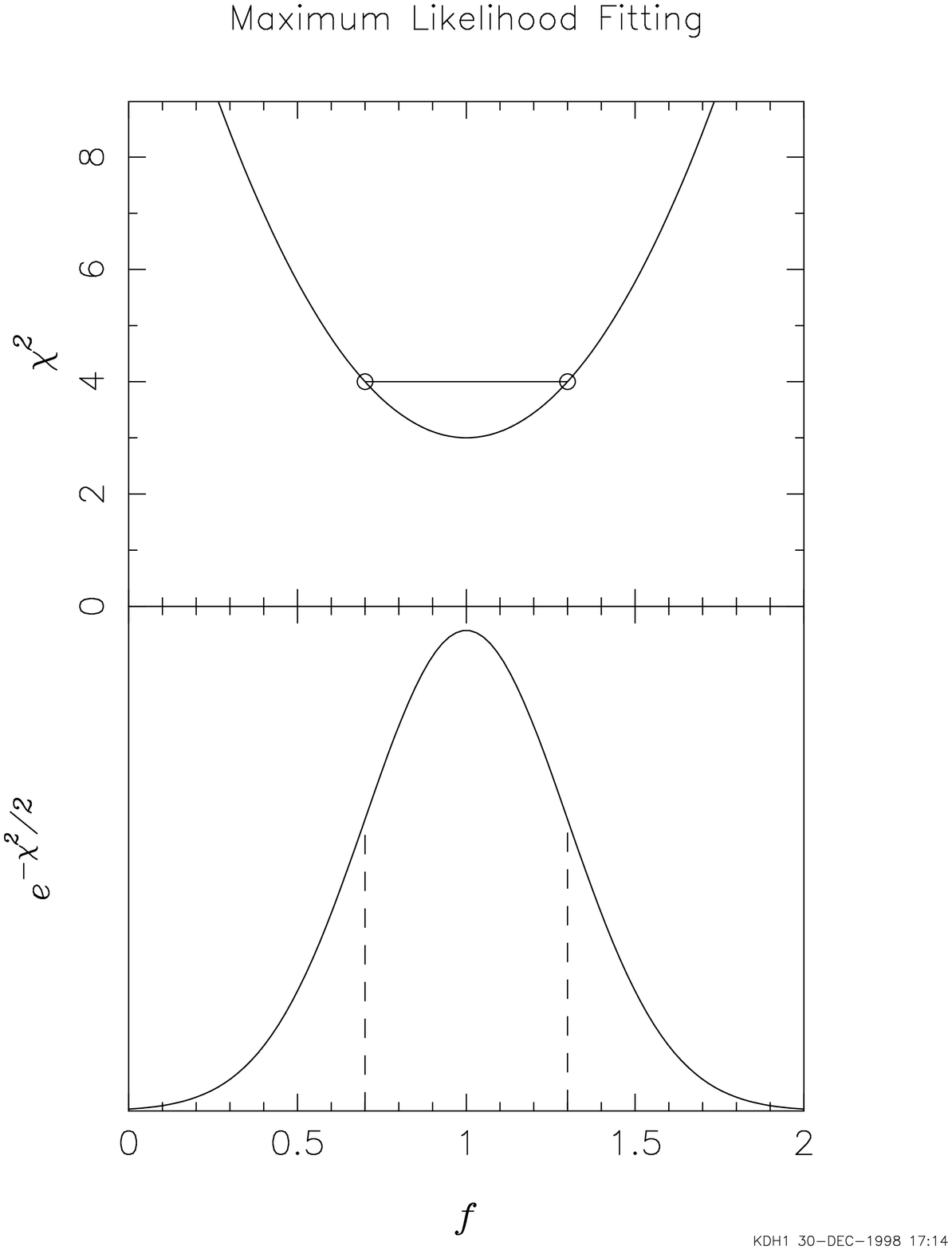}}
\end{picture}\par
\caption[]
{\small
The $\pm\sigma$ confidence interval and the 
posterior probability distribution
are shown for one parameter of a maximum likelihood fit.
\label{fig:mlf}}
\end{figure}

The maximum likelihood method 
assumes that $P(\vec{f})$ is a uniform distribution.
The ``best fit'' model $\hat{f}$ is the one that maximizes the likelihood.
This is equivalent (when $\sigma_i$ are known) to minimizing $\chi^2$.
A model using $M$ parameters to fit $N$ data points should leave
residuals with $N-M$ degrees of freedom.
A successful fit should therefore achieve 
$\chi^2_{\sc min}/(N-M) \approx 1 \pm \sqrt{2/(N-M)}$.
The $\pm\sigma$ interval for any parameter of the fit can
be found by the criterion $\chi^2 < \chi^2_{\sc min} + 1$.
These concepts are illustrated in Fig.~\ref{fig:mlf}.

\begin{figure}
\vspace*{10.5cm}
\begin{picture}(8,11)
\epsfxsize=8cm
\epsfysize=11cm
 \makebox[8cm][l]{\epsfbox{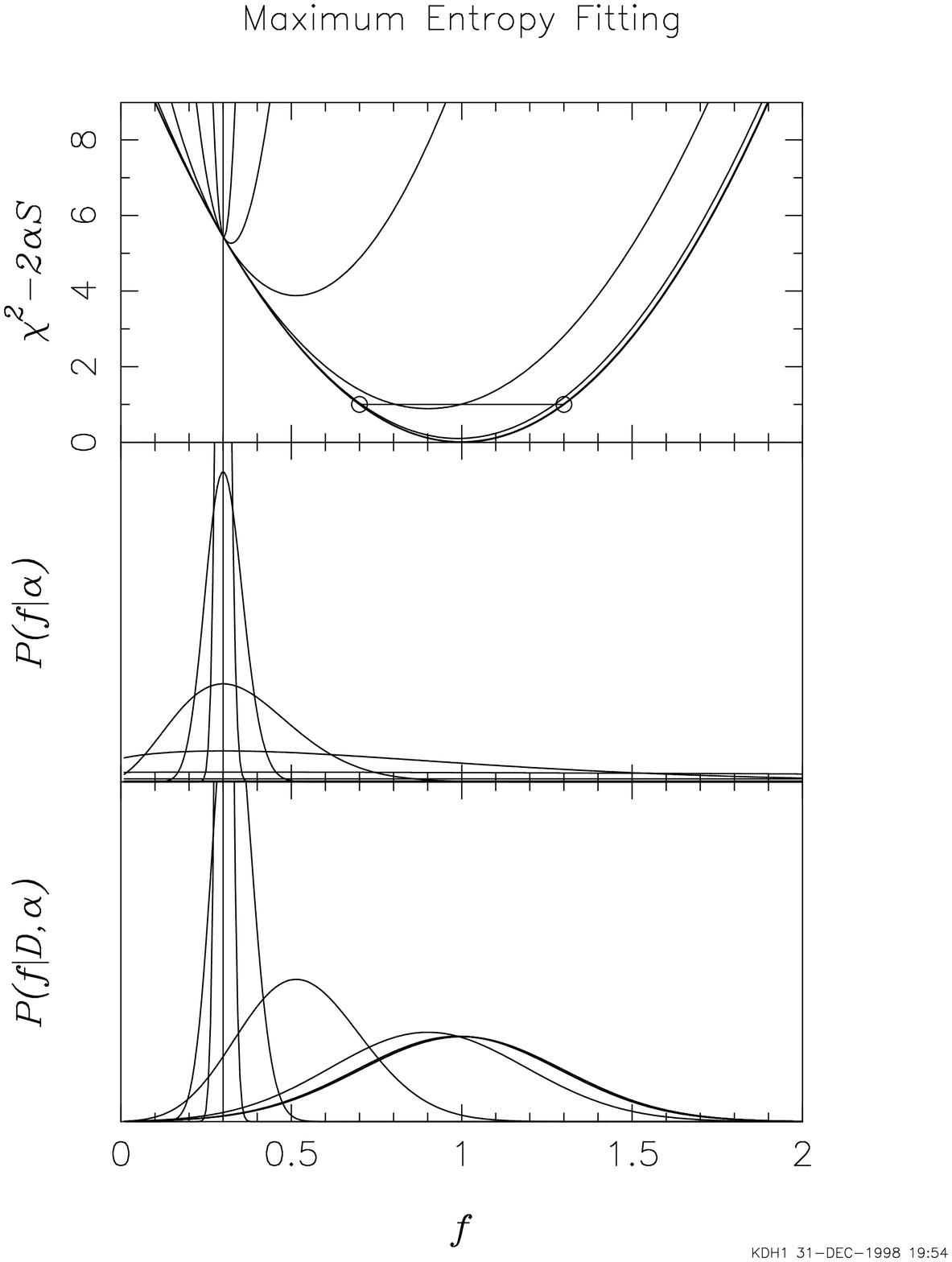}}
\end{picture}\par
\caption[]
{\small
In maximum entropy fitting,
we minimize $\chi^2 - 2 \alpha S$.
The entropic prior $P(f|\alpha) \propto {\rm exp}\left\{\alpha S \right\}$
peaks at the default value $d$ and has an rms width $(d/\alpha)^{1/2}$.
When the regularization parameter $\alpha$ is very small,
the prior is so wide that the entropy $S$ is ignored and the
posterior probability becomes
$P(f|D,\alpha) \propto {\rm exp}\left\{-\chi^2/2 \right\}$.
As $\alpha$ increases, the entropy becomes more important.
$P(f|D,\alpha)$ then shifts toward the default value and becomes narrower.
The optimum value of $\alpha$ is determined in Fig.~\ref{fig:alp}.
\label{fig:mef}}
\end{figure}

The Bayesian formulation includes the prior $P(\vec{f})$
to make explicit a fundamental ambiguity involved
in the interpretation of data.
Different people may use different priors, and thereby
arrive at different conclusions from the same data.
This is illustrated in Fig.~\ref{fig:mef}.

\subsection{ For a Simple Map: Maximize Entropy }

The maximum entropy method extends maximum likelihood fitting
to include models that employ huge numbers of parameters.
The entropy quantifies the number of parameters being used
in such fits.
Maximum entropy fitting seeks the simplest model
that fits the data.

In quasar tomography,
our $M$-dimensional parameter vector $\vec{f}$
(equivalently, $M$-pixel map)
arises from partitioning the domains of several
{\it positive additive distributions},
e.g., $f(R,\theta,n_{\sc H},N_{\sc H},v)$, $L(t)$, 
$F_B(\lambda)$, and $D$.
If we wish to ensure that our maps take on only positive values,
and that our inferences depend on the underlying distributions
represented by the maps, and not on specific coordinates
or partitions used to slice up the domain of $\vec{f}$,
then the only consistent way to do this (Skilling 1989)
is to assign relative probabilities to the 
various possible maps $\vec{f}$ 
according to an {\it entropy} of the form
\begin{equation}
S(\vec{f})
 = \sum_{i=1}^M w_i 
\left\{ f_i - d_i - f_i \ln{(f_i/d_i)} \right\}\ .
\end{equation}
Here $w_i$ is a weight proportional to the {\it volume} of pixel $i$,
and $d_i(\vec{f})$ is the {\it default value} of $f_i$.

The entropy is regarded as measuring the ``simplicity'' of the map.
The simplest map maximizes the entropy, and therefore satisfies
\begin{equation}
0 = \frac{\partial S}{\partial f_i}
= - w_i \ln{ ( f_i / d_i ) } 
+ \sum_{k=1}^M w_k \left( \frac{f_k}{d_k} - 1 \right) 
\frac{\partial d_k}{\partial f_i}\ .
\end{equation}
Note that the maximum entropy value, $S=0$,
occurs when $f_i = d_i$ for every pixel $i$.
The entropy is an information-like measure of the distance
of $\vec{f}$ from its {\it default map} $\vec{d}(\vec{f})$.
Maximizing the entropy ``steers'' $\vec{f}$
toward $\vec{d}$, and that is why
the $d_i$ are referred to as default values.
We discuss in \S\ref{sec:def} our prescription for
chosing the default map $\vec{d}(\vec{f})$
to make the entropy express a preference for specific
symmetries, for example smoothness, or axi-symmetry.

\subsection{ The Entropic Prior }

The prior probability associated with the entropy
(Skilling 1989) is 
\begin{equation}
	P(\vec{f}|\alpha) = 
	\prod_{i=1}^M P(f_i|\alpha) = 
	\frac{ {\rm exp}\left\{ \alpha S(\vec{f}) \right\} } { Z_S(\alpha) }\ .
\end{equation}
This introduces a {\it regularization parameter}, $\alpha > 0$, 
and the corresponding partition function,
\begin{equation}
Z_S(\alpha) = \int P(\vec{f}|\alpha)\ d\vec{f}
	= \int {\rm exp}\left\{ \alpha S(\vec{f}) \right\}\ d\vec{f}\ ,
\end{equation}
which normalizes the entropic prior to a probability density
on $\vec{f}$.

To gain some insight into the meaning of this expression,
expand $S(\vec{f})$ in a Taylor series
about its peak at $\vec{f} = \vec{d}$, and truncate this to obtain
the quadratic approximation
\begin{equation}
	S(\vec{f}) \approx \frac{1}{2}
	\sum_{i,j=1}^M \left( f_i - d_i \right)
	\frac{\partial^2 S(\vec{d})}{\partial f_i \partial f_j}
	\left( f_j - d_j \right)\ .
\end{equation}
The corresponding entropic prior is an $M$-dimensional Gaussian,
\begin{equation}
P(\vec{f}|\alpha) \approx \frac{
	{\rm exp}\left\{ \frac{\alpha}{2} \sum_{i,j}
	\left( f_i - d_i \right)
	\frac{\partial^2 S(\vec{d})}{\partial f_i \partial f_j}
	\left( f_j - d_j \right) \right\}
	} { Z_S(\alpha) }\ .
\end{equation}
The entropy curvature matrix is given in general by
\begin{equation}
\begin{array}{ll}
\frac{\partial^2 S(\vec{f})}{\partial f_i \partial f_j } 
& = - \frac{w_i}{f_i} \delta_{ij}
 + \frac{w_i}{d_i} \frac{\partial d_i}{\partial f_j}
 + \frac{w_j}{d_j} \frac{\partial d_j}{\partial f_i}
\\ & 
+ \sum_{k=1}^M
w_k \left\{
\left( \frac{f_k}{d_k} - 1 \right)
	\frac{\partial^2 d_k}{\partial f_i \partial f_j}
-\frac{f_k}{d_k^2}
	\frac{\partial d_k}{\partial f_i}
	\frac{\partial d_k}{\partial f_j}
\right\}\ .
\end{array} 
\end{equation}
For our present purposes, specialize to the case
of a fixed default map, $\partial d_i / \partial f_j \ll 1$,
for which the entropy curvature matrix is diagonal,
\begin{equation}
\frac{\partial^2 S(\vec{f})}{\partial f_i \partial f_j } 
	\approx - \frac{w_i}{f_i} \delta_{ij}
\ .
\end{equation}
The entropic prior is then a product of $M$ Gaussian distributions
\begin{equation}
P(\vec{f}|\alpha) \approx \frac{
{\rm exp}\left\{ - \frac{ \alpha}{2} \sum_{i=1}^{M}
	\frac{ w_i }{ d_i } \left( f_i - d_i \right)^2 \right\}
}{Z_S(\alpha)}
\ ,
\end{equation}
with the partition function
\begin{equation}
Z_S(\alpha) \approx 
	\prod_{i=1}^M \left( \frac{2 \pi d_i}{\alpha w_i} \right)^{1/2}
\ .
\end{equation}
From this we see that the entropic prior admits values in the range
$f_i \sim d_i \left( 1 \pm \left( \alpha w_i d_i \right)^{-1/2} \right)$.
For $\alpha \gg ( w_i d_i )^{-1} $, the prior
confines attention to a narrow range $f_i \sim d_i$,
while for $\alpha \ll ( w_i d_i )^{-1}$, a wider range
opens to consideration (see Fig.~\ref{fig:mef}).
In this way $\alpha$ controls the strength
of our prior conviction that the default values $d_i$ are correct.
If we wish to set $\alpha$ to 
a different value for each pixel,
this can be done effectively 
by changing the pixel weights $w_i$.

\subsection{ The Maximum Entropy Trajectory }

Given a dataset $\vec{D}$, and a regularization parameter $\alpha$,
the posterior probability of our map $\vec{f}$ is
\begin{equation}
\label{eqn:P(f|D,a)}
P(\vec{f}|\vec{D},\alpha) 
	= \frac{
	{\rm exp}\left\{ \alpha S - \chi^2/2 \right\}
		}{ Z_Q(\alpha) }\ ,
\end{equation}
where the appropriate partition function is
\begin{equation}
	Z_Q(\alpha) =
\int {\rm exp}\left\{ \alpha S - \chi^2/2 \right\}\ d\vec{f}\ .
\end{equation}
As $\alpha \rightarrow \infty$, the entropy term dominates and the
prior becomes so narrow that it over-rules any influence of the data.
In the opposite limit, $\alpha \rightarrow 0$,
the prior becomes very wide, and we revert to the maximum likelihood
fit minimizing $\chi^2$.
Thus $\alpha$ parameterizes a 1-dimensional family of maps $\hat{f}(\alpha)$,
{\it the maximum entropy trajectory} (Gull \& Skilling 1990),
connecting the maximum likelihood map ($\alpha=0$),
and the maximum entropy map ($\alpha \rightarrow \infty$).

To explore this family of maximum entropy fits,
we use the MEMSYS algorithm (Skilling \& Bryan 1984)
to make iterative adjustments to $\vec{f}$
in order to maximize
\begin{equation}
	Q = \alpha S\ - \chi^2/2 .
\end{equation}
For any given value of $\alpha$,
the iteration proceeds until the gradients of $\chi^2$ and
$S$ are found to be parallel to machine precision,
thus assuring that $Q$ is maximized.
We begin where we must, with a high value of $\alpha$,
and a simple map that fits the data poorly.
We then gradually lower $\alpha$ to reduce the
influence of the entropy in comparison with that of the data.
For each lower value of $\alpha$, the map develops
additional structure in order to improve the fit.
Thus $\chi^2$ decreases monotonically as we pass along the
maximum entropy trajectory through 
progressively lower values of $\alpha$.

\subsection{ Stopping Criteria }

Which value of $\alpha$ should we choose?
How strong is our faith that the default values are correct?
A justifiable strategy rejects high values of $\alpha$ that
produce poor fits to the data, and low values of $\alpha$
that fit the data too well.
Based on this motivation, a common practice is to select
a value of $\alpha$ that achieves $\chi^2/N = 1$.
This is the stopping criterion we have adopted for
all the results shown in this paper.

The $\chi^2=N$ stopping criterion gives a conservative fit.
It under-fits the data a bit and omits some of the finer-scale
structure in the map.
To see this, consider a maximum likelihood fit with $M$ parameters
adjusted to fit $N$ data points.
The best fit is expected to leave residuals with $N-M$ degrees of freedom,
i.e. $\chi^2 \sim (N-M) \pm \sqrt{ 2(N-M) }$.
A fit that achieves $\chi^2 \sim N$ therefore under-fits the data.

\begin{figure}
\vspace*{10.5cm}
\begin{picture}(8,11)
\epsfxsize=8cm
\epsfysize=11cm
 \makebox[8cm][l]{\epsfbox{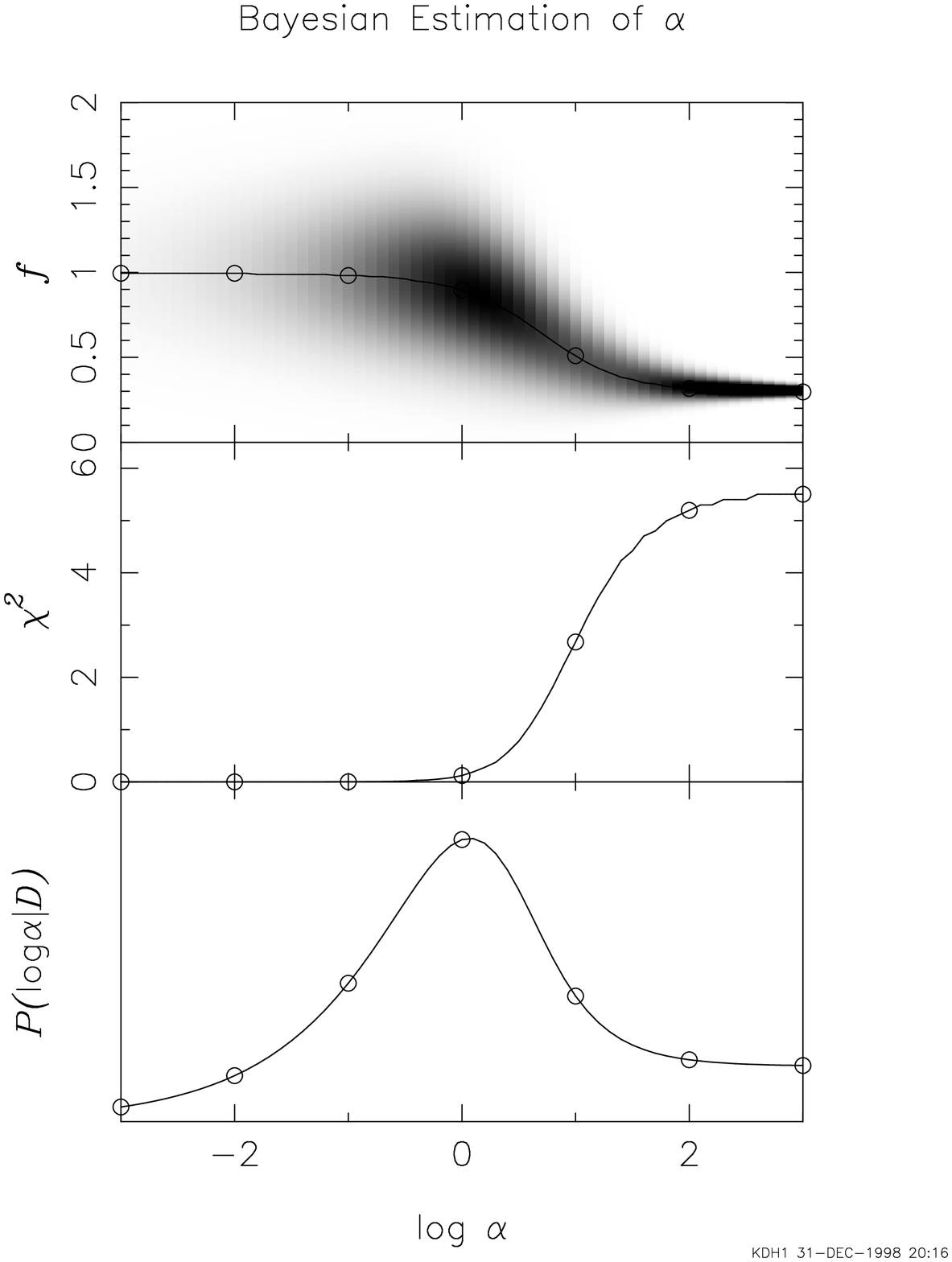}}
\end{picture}\par
\caption[]
{\small
Bayesian determination of the regularization parameter $\alpha$
for the problem illustrated in Fig.~\ref{fig:mef}.
If $\alpha$ is large, the entropy pulls the solution to the default
value, and the probability is then low because the fit
to the data is poor.
If $\alpha$ is small, the entropy is ignored.
The fit is then good, but the probability is low
because there is a penalty for using a very wide prior.
The probability peak gives the most likely value of $\alpha$.
\label{fig:alp}}
\end{figure}

A Bayesian stopping criterion aims to maximize
the posterior probability of $\alpha$,
given by
\begin{equation}
\begin{array}{rl}
P(\alpha|\vec{D}) & \propto\ P(\vec{D}|\alpha)\ P(\alpha)
\\ & =\ P(\alpha)\ \int P(\vec{D}|\vec{f})\ P(\vec{f}|\alpha)\ d\vec{f}
\\ & =\ P(\alpha)\ \frac{ Z_Q(\alpha) } {Z_D\ Z_S(\alpha) }\ .
\end{array}
\end{equation}
This entails some ambiguity because we are free
to choose the prior $P(\alpha)$.
Since $\alpha>0$, a plausible choice is $P(\alpha) \propto \alpha^{-1}$,
so that the prior uniform in $\ln{\alpha}$.
For an example, see Fig.~\ref{fig:alp}.

We expect $P(\alpha|\vec{D})$ to have a
maximum at some $\alpha = \hat{\alpha}$,
provided $P(\alpha)$ is not too bizarre.
To see this, note that for $\alpha \gg \hat{\alpha}$
the emphasis on entropy maximization makes the map very rigid.
$P(\alpha|\vec{D})$ will then be small because $\chi^2$ is large
-- very simple maps fail to achieve good fits to the data.
As we decrease $\alpha$, we move along the maximum entropy trajectory
through increasingly flexible maps that develop more structure
to improve the fit.
For $\alpha \ll \hat{\alpha}$, the map is so flexible
that $\chi^2$ approaches an asymptotic minimum value
as the fit becomes as good as it can be.
In this limit $P(\alpha|\vec{D}) \rightarrow 0$
because $Z_S(\alpha) \propto \alpha^{-M/2}$.
The entropic prior is imposing an appropriate penalty (Occam's razor)
on very flexible maps that deploy too many degrees of freedom
to achieve their fit.
For $\alpha = \hat{\alpha}$ we achieve the best compromise --
fitting the data well enough while deploying a minimum
number of degrees of freedom in the map.

How many parameters do we employ in a regularized fit?
The map may have an enormous number of pixels --
in fact $M$ may be larger than $N$ --
so we can't possibly be using all $M$ degrees of freedom.
In fact we are not, because the structure that develops in the map
is spatially correlated.
Because we are minimizing $\chi^2 - 2 \alpha S$,
it is plausible to identify $- 2 \alpha S$ as the 
effective number of parameters used in the regularized fit.
(Note that $-2\alpha S>0$ because $\alpha>0$ and $S<0$.)
By analogy with a maximum likelihood fit, then,
we should choose $\alpha$ so that $\chi^2 = N + 2 \alpha S$.
This is only a heuristic justification, but
the result can be justified more rigorously (Gull 1989)
when the Bayesian stopping criterion is used.

\subsection{Quantified Uncertainties}

The uncertainty of the map $\vec{f}$ after fitting to the data $\vec{D}$
is quantified by its posterior probability distribution.
Equation~(\ref{eqn:P(f|D,a)}) gives this for each value of $\alpha$,
corresponding to different strengths of faith in the default map.

One can combine these by performing a Bayesian average over $\alpha$
to obtain
\begin{equation}
\begin{array}{rl}
P(\vec{f} | \vec{D} ) 
	& = \int P(\vec{f} | \vec{D}, \alpha)\
	P(\alpha|\vec{D})\ d\alpha
\\ & \propto \int 
	\frac{{\rm exp}\left\{\alpha S - \chi^2/2 \right\}}
	{Z_D\ Z_S(\alpha)}\
	P(\alpha)\ d\alpha\ .
\end{array}
\end{equation}
In many cases, particularly with large numbers of
data points and map pixels,
the probability peak is narrow enough to permit the use of
quadratic approximations to $\chi^2$ and $S$,
yielding an M-dimensional Gaussian probability density
centred on the most probable map on the maximum entropy 
trajectory at $\hat{f}(\hat{\alpha})$.

Inferences about other parameters of the fit,
such as the distance, luminosity, and $H_0$,
may be quantified in the same way,
\begin{equation}
\begin{array}{rl}
P( H_0 | \vec{D} ) 
	& \propto P(\vec{D}|H_0)\ P(H_0)
\\	& = P(H_0)\ \int P(\vec{D} | H_0, \vec{f} )\
	P(\vec{f}|\alpha)\ d\vec{f}\ P(\alpha) d\alpha
\\ & = P(H_0)
	\int \frac{ {\rm exp}\left\{\alpha S - \chi^2/2 \right\} }
		{ Z_D\ Z_S(\alpha) }\ d\vec{f}\
	P(\alpha)\ d\alpha\ 
\\ & = P(H_0)
	\int \frac{ Z_Q(\alpha) } { Z_D\ Z_S(\alpha) }\
	P(\alpha)\ d\alpha\ .
\end{array}
\end{equation}

\subsection{ Default Maps }
\label{sec:def}

How shall we choose defaults?
We want the entropy to ``steer'' the solution toward
the ``simplest'' map that fits the data.
The default map should therefore be set to whatever we
consider to be the simplest map.
But what do we consider be a simple map?

In general, we compute default values as weighted geometric means
of ``nearby'' map values:
\begin{equation}
d_i(\vec{f}) = {\rm exp}\left\{ \sum_{j=1}^M B_{ij} \ln{f_j} \right\}
\ .
\end{equation}
Think of the symmetric ``blur'' matrix $B_{ij}$ as a point-spread
function, decreasing with the ``distance'' between pixels $i$ and $j$.
It is normalized as a probability distribution
\begin{equation}
\sum_{j=1}^M B_{ij} = 1
\ .
\end{equation}
This makes $\vec{d}(\vec{f})$ a ``blurred'' copy of $\vec{f}$.
Each pixel in $\vec{f}$ is then pulled toward its neighbors.

If we set $B_{ij}=1/M$,  then the default map is uniform,
and the maximum entropy fit gives the ``most uniform'' map that
fits the data.  This would be appropriate when the prior
expectation is that no structure is present in $\vec{f}$.
In problems with incomplete data constraints,
a uniform default can allow the data to impress
bizarre artifacts on the map,
for example the ingress-egress arches seen in
eclipse maps (Horne 1985).

It is usually more appropriate to seek smooth maps
by using ``curvature defaults'' that steer each pixel value
toward the geometric mean of its nearest neighbors,
\begin{equation}
d_i = \sqrt{ f_{i+1} f_{i-1} }\ .
\end{equation}
With curvature defaults,
the entropy of a 1-dimensional map $f(x)$,
partitioned at intervals $\Delta x$,
approaches a weighted integral of the squared curvature of
the logarithm of the map,
\begin{equation}
S \approx - \int \frac{ (\Delta x)^4 w(x) f(x) }{2}
\left( \frac{\partial^2 \ln{f}}{\partial x^2} \right)^2\ .
\end{equation}
With constant $w(x)$, this gives preference
to $f(x)$ with Gaussian peaks (constant curvature)
and exponential tails (zero curvature).
With $w(x) \propto x^2$,
the solution is steered toward $f(x)$ that are power-laws in $x$.

In dealing with multi-dimensional maps, for example
$f(\vec{x}) = f(R,\theta,n_{\sc H},N_{\sc H},v)$,
there are of course many different curvatures to consider.
We compute the default value for pixel $\vec{x}$
by first calculating geometric means of neighbors
offset on opposite sides in several chosen directions $\Delta\vec{x}_k$,
and then taking a weighted average of those geometric means:
\begin{equation}
\ln{d(\vec{x})} = 
\frac{ 
\sum_k \left( \Delta_k / \Delta\vec{x}_k \right)^2
\ln{ \sqrt{ f(\vec{x}+\Delta\vec{x}_k) 
	f(\vec{x}-\Delta\vec{x}_k) } }
}{ 
\sum_k \left( \Delta_k / \Delta\vec{x}_k \right)^2
}\ .
\end{equation}
A large $\Delta_k$ promotes
structures oriented along the direction $\Delta\vec{x}_k$.

We frequently use the default map to steer our solutions 
toward certain preferred symmetries.
For example, to promote a spherical geometry, 
we would compute defaults by averaging map values 
over spherical shells.
Point-symmetric and axi-symmetric geometries
project to maps with front-to-back symmetry,
i.e. symmetric on reflection through the $Y$ axis.  
To promote point- or axi-symmetric geometries,
we steer each pixel toward the mean of itself and
the pixel with opposite sign of $\cos{\theta}$.
We also apply constraints on the normalization of the map,
for example covering fraction must be less than 100\%,
by imposing the required constraint on the default map.
In this way the desired symmetry and normalization are
taken on by the map if this is compatible with the data.
Otherwise the map comes as close as it can to the desired
symmetry and normalization while still fitting the data.

Finally, we note that it may be possible by making a deft choice 
of default image to suppress to some extent
the artefacts that form along the time-delay paraboloids.
The strategy would be to first blur the map along the time delay direction,
and then mix that in with a negative weight to form the final default image.
With negative weight the default will suppress rather than enhance
structure in the time-delay direction.
This apporach has demonstrated some success in suppressing
similar artefacts in the eclipse mapping problem (Spruit 1994).
A difficulty we see with this approach is in deciding how
strongly to apply the negative weights.
We have not attempted to explore this possibility, but may
do so in future work.

\label{lastpage} 

\end{document}